\documentclass[aps, prb, twocolumn, superscriptaddress, longbibliography]{revtex4-2}

\usepackage{mathrsfs}
\usepackage{tabularx}
\usepackage{slashed}
\usepackage{amsmath}
\usepackage{amsfonts}
\usepackage{graphicx,color}
\usepackage{times}
\usepackage[caption=false]{subfig}
\usepackage[colorlinks,linkcolor=red,citecolor=blue]{hyperref}
\newcommand{\comment}[1]{}
\AtBeginDocument{%
    \newwrite\bibnotes
    \def\bibnotesext{Notes.bib}
    \immediate\openout\bibnotes=\jobname\bibnotesext
    \immediate\write\bibnotes{@CONTROL{REVTEX42Control}}
    \immediate\write\bibnotes{@CONTROL{%
    apsrev42Control,author="08",editor="1",pages="1",title="0",year="1"}}
     \if@filesw
     \immediate\write\@auxout{\string\citation{apsrev42Control}}%
    \fi
}%
\usepackage[english]{babel}
\usepackage{amsmath,amssymb,bbm,mathrsfs,bm,braket,color,graphicx,comment,amsfonts,dsfont}
\usepackage{soul}
\usepackage{xcolor}
\newcommand{\pd}{{\phantom{\dag}}}
\newcommand{\abs}[1]{\left|#1\right|}

\begin{document}
\newcommand{\janos}[2]{\st{#1}{\color{olive} {#2}}}
\newcommand{\hui}[2]{\st{#1}{\color{blue} {#2}}}
\newcommand{\emil}[2]{\st{#1}{\color{brown} {#2}}}

\title{Topological fine structure of an energy band}

\author{Hui Liu}
\affiliation{Department of Physics, Stockholm University, AlbaNova University Center, 106 91 Stockholm, Sweden}

\author{Cosma Fulga}
\affiliation{Leibniz Institute for Solid State and Materials Research,
IFW Dresden, Helmholtzstrasse 20, 01069 Dresden, Germany}
\affiliation{W\"{u}rzburg-Dresden Cluster of Excellence ct.qmat, 01062 Dresden, Germany}

\author{Emil J. Bergholtz}
\affiliation{Department of Physics, Stockholm University, AlbaNova University Center, 106 91 Stockholm, Sweden}

\author{J\'anos K. Asb\'oth}
\affiliation{Department of Theoretical Physics, Institute of Physics, 
Budapest University of Technology and Economics, M\H uegyetem rkp. 3., H-1111 Budapest, Hungary}
\affiliation{HUN-REN Wigner Research Centre for Physics, H-1525 Budapest, P.O. Box 49., Hungary}
%Janos2026
\affiliation{HUN-REN-BME-BCE Quantum Technology Research Group, Budapest University of Technology and Economics, M\H uegyetem rkp. 3., H-1111 Budapest, Hungary}

\begin{abstract}
A band with a nonzero Chern number cannot be fully localized by weak disorder.
There must remain at least one extended state, which ``carries the Chern number.''
Here we show that a trivial band can behave in a similar way.
Instead of fully localizing, arbitrarily weak disorder leads to the emergence of two sets of extended states, positioned at two different energy intervals, which carry opposite Chern numbers.
Thus, a single trivial band can show the same behavior as two separate Chern bands.
We show that this property is predicted by a topological invariant called a ``localizer index.''
Even though the band as a whole is trivial as far as the Chern number is concerned, the localizer index allows access to a topological fine structure.
This index changes as a function of energy within the bandwidth of the trivial band, causing nontrivial extended states to appear as soon as disorder is introduced.
Our work points to a previously overlooked manifestation of topology, which impacts the response of systems to impurities beyond the information included in conventional topological invariants.
\end{abstract}

\maketitle

\emph{Introduction} --- 
Anderson localization ~\cite{Anderson, evers_mirlin} describes how disorder localizes quantum states, causing wavefunctions to acquire finite localization lengths and turning a conductor into an insulator.
Initially, the scaling theory of localization~\cite{scaling_theory, Wegner1979} predicted that in two-dimensional (2D) systems without symmetry, i.e., class A of the Altland-Zirnbauer classification~\cite{tenfold_symmetry}, only insulating phases exist: all states of a thermodynamically large system are localized for arbitrarily weak disorder.

The discovery of the quantum Hall effect~\cite{qhe_rmp}, and the subsequent development of the theory of topological phases of matter~\cite{qi_zhang_rmp, Hasan_Kane_rmp, Chiu_rmp}, changed this paradigm by showing that not all 2D class A insulators are topologically equivalent. 
This allows for insulator-to-insulator transitions between distinct topological phases. In the thermodynamic limit, these transitions are marked by robust, critically delocalized states that reside at isolated energies.
Consequently, a band with nonzero Chern number cannot be fully localized by weak disorder~\cite{Laughlin}.
Instead, at least one extended state that ``carries the Chern number'' must remain \cite{Halperin}.
Using terminology introduced by Laughlin~\cite{Laughlin}, extended states at different isolated energies and carrying opposite Chern numbers ``levitate'' towards each other, eventually ``annihilating'' in order to produce a trivial Anderson insulator.

In contrast, trivial bands, i.e., those which are not characterized by any nonzero Chern numbers, are generically expected to fully localize even for infinitesimal disorder strength, provided that the disorder is sufficiently generic~\cite{evers_mirlin, Huckestein_rmp, IQHE_lattice,Prodan_2010,Onoda_qshe,3dTI,Titum_afai, Roy_2016, Liu_2020,PhysRevB.109.125132}, see Fig.~1(a). 
Exceptions include systems with nonzero mirror- or spin-Chern numbers, such as $\mathbb{Z}_2$ topological insulators with spin-rotation symmetry, where spin-polarized critical states carrying opposite Chern numbers can remain delocalized under weak disorder~\cite{PhysRevB.85.075115}.

\begin{figure}
\centering
\includegraphics[width=\linewidth]{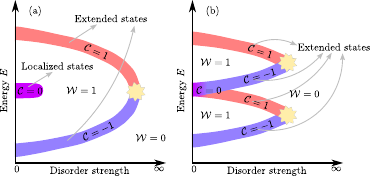}
\caption{
Panel (a): The paradigm of Anderson localization for 2D systems in class A.
A trivial band (middle, Chern number ${\cal C}=0$) is fully localized by disorder, whereas nontrivial bands (${\cal C}=\pm 1$) lead to robust extended states.
The latter eventually meet and annihilate (star) for larger disorder strength.
The index ${\cal W}$ denotes the number of chiral edge modes present at energies between/outside those of the extended states.
Panel (b): Our main result.
The trivial band does not fully localize, but splits into two branches of extended states carrying opposite Chern numbers. As we show, this behavior is a consequence of its topological fine structure.
\label{fig:main_idea}
}
\end{figure}

Here, we revisit Anderson localization of trivial bands of 2D class A systems. 
We demonstrate that vanishing Chern numbers do not preclude the existence of multiple, robust extended states, even in the absence of internal degrees of freedom or hidden symmetries.
Instead, there are multiple, robust extended states, which carry opposite Chern numbers, and which, as disorder is further increased,  participate in the levitation and annihilation process [see Fig.~\ref{fig:main_idea}(b)].

We characterize this behavior via the \emph{spectral localizer}, which provides an energy-resolved distribution of topological charge~\cite{spectral_localizer, LORING2015383, Loring_Hermann_2, Loring_finite_volume}, referred to as the \emph{topological fine structure}.
While the Chern number is a global property characterizing the entirety of a band, the localizer index can be evaluated at a given energy, thus providing an internal topological structure.
As we show, changes of this topological invariant necessarily lead to the formation of critical states that are robust against disorder.

\begin{figure}
\centering
\includegraphics[width=\linewidth]{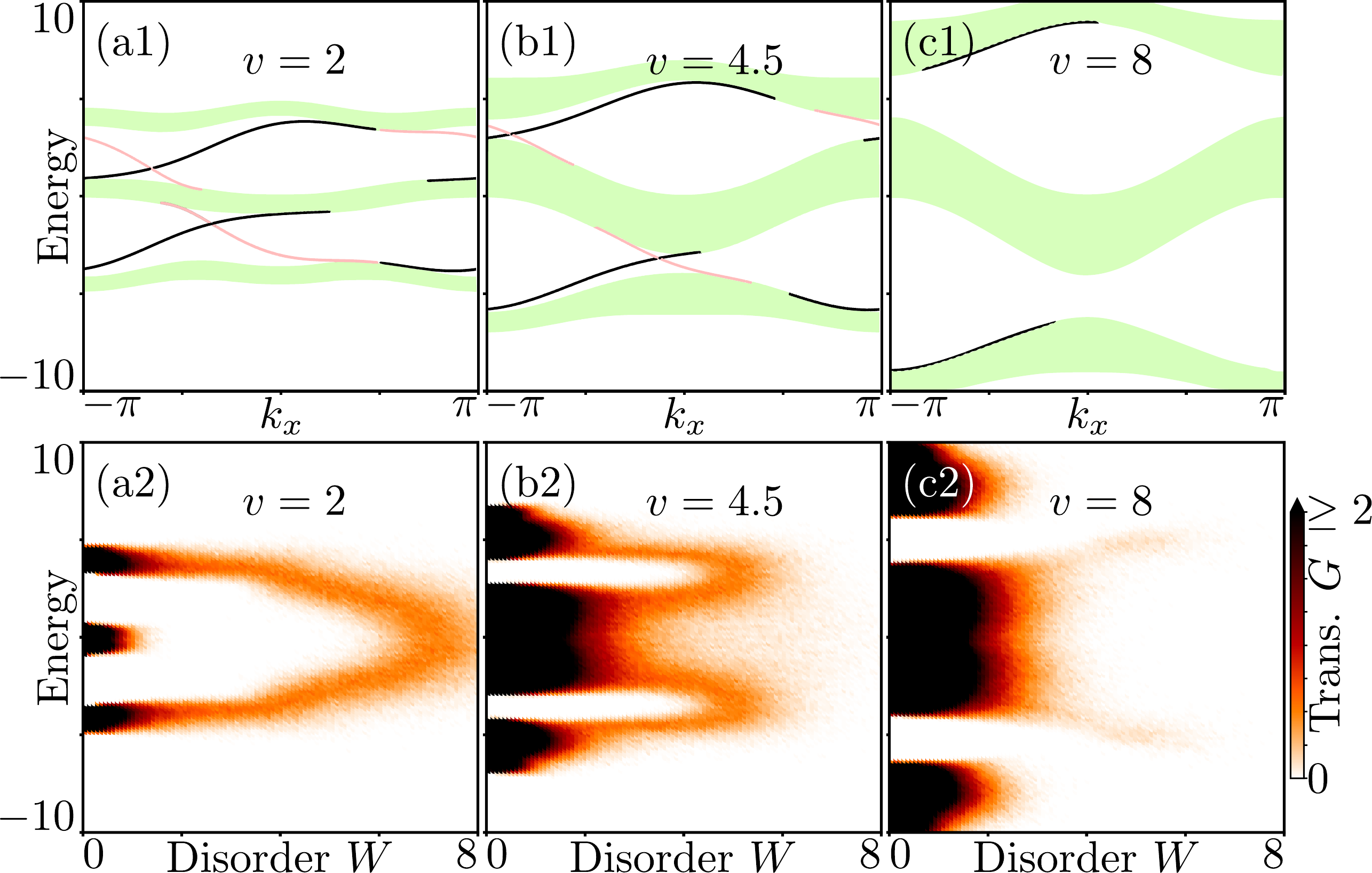}
\caption{
Top panels: band structure of the model in the absence of disorder. We use a ribbon geometry, infinite along the $x$-direction and consisting of 60 unit cells along $y$. Here, we use $p_c=\sum_{i=1}^{N_y/2}|\psi_i|^2$ to label the color, where $0<p_c<0.49$ (black), $0.51\leq p_c\leq 1$ (red), and $0.49\leq p_c\leq 0.51$ (green) indicate states localized at the top edge, bottom edge, and in the bulk, respectively.
Bottom panels:
Two-terminal transmission probability, plotted as a function of energy $E$ and disorder strength $W$, for a $60\times60$ unit cell system with periodic boundary conditions in the $y$-direction.
Darker regions correspond to the presence of states extended on the scale of the finite-size sample, helping us track the levitation and annihilation process. Each point is obtained by averaging over 50 independent disorder realizations. \label{fig:evolution}
}
\end{figure}

\emph{An example} --- We begin by illustrating our general conclusions using a simple 2D model: a two-band Chern insulator coupled to a single, trivial band.
The momentum-space Hamiltonian reads
\begin{equation}
    H(\mathbf{k})=\begin{pmatrix}
        h^\pd_{11}(\mathbf{k})&h^\pd_{12}(\mathbf{k})&v\\
        h_{12}^{*}(\mathbf{k})& -h^\pd_{11}(\mathbf{k})&0\\
        v&0&0
    \end{pmatrix},
    \label{eq:Hamiltonian}
\end{equation}
where $\mathbf{k}=(k_x, k_y)$ is the 2D quasimomentum, and the matrix element functions are $h_{11}(\mathbf{k})=2(\cos k_x - \cos k_y)$, and $h_{12}(\mathbf{k})=\sqrt{2}e^{-i\pi/4}(e^{ik_x}+e^{ik_y}+ie^{i(k_x+k_y)}+1)$.
The upper-left $2\times 2$ block is a Chern insulator with Chern numbers $\mathcal{C}=\pm1$ for the upper and lower bands~\cite{fqhe_zero_mag}, whose hopping amplitudes are used to set the unit of energy. 
The lower right element is the single-band atomic insulator: a trivial flat band with a vanishing onsite potential. It is coupled to the Chern insulator with an amplitude $v$ (see the real space lattice and its corresponding Hamiltonian in the Appendix~\ref{sec:real space lattice}).  

In the regime of coupling strengths $0\le v <8$, this three-band model has two phases: a topological phase for small coupling, $0\leq v \lesssim 5.65$, and a trivial phase for large coupling, $v \gtrsim 5.65$. In the topological phase the top and bottom bands have Chern numbers $+1$ and $-1$, respectively, whereas the middle band is trivial, with Chern number 0. In the trivial phase all bands have a vanishing Chern number. These two phases are separated by a topological phase transition, where the gaps between the 1st and 2nd, and the 2nd and 3rd band close simultaneously [see Fig.~\ref{fig:evolution}(a1-c1)].

We are interested in what happens to the energy eigenstates of the model as disorder is turned on and gradually increased. Thus, we add a random onsite potential to the Hamiltonian, different for each of the three orbitals, uniformly distributed in $[-W, W]$, with $W$ denoting the disorder strength. 
We calculate the two-terminal transmission probability $G$ of a finite-size, square system using the Kwant package~\cite{Groth_2014} in order to numerically estimate the energies at which extended states are present.
The leads are attached to the left and right boundaries of the system, and we connect the top and bottom by periodic boundary conditions along the transverse direction, so as to pick up only the contribution of bulk states to the transmission probability.
For additional details of the numerical implementation, see the Appendix~\ref{sec:transport setting} and the code on Zenodo~\cite{zenodo}. 

Localized states' contribution to the transmission probability decays exponentially with system size, whereas for extended bulk states this contribution stays constant, or even grows, as system size is increased.
Thus, for a sufficiently large system (in our numerics, $60\times60$ unit cells), we can use $G(E, W)$ to locate energies where extended states exist [large transmission probabilities, darker colors in Fig.~\ref{fig:evolution}(a2-c2)], and then track the pattern of their levitation and annihilation. Note that in these finite size calculations, localized states with localization length comparable to or larger than the system size also contribute to the conductance, giving significant finite-size conductance values at weak disorder. 

When the atomic insulator and Chern insulator are weakly coupled, $v \lesssim 3$, the pattern of levitation and annihilation is unsurprising. 
As seen in Fig.~\ref{fig:evolution}(a2), first the middle, trivial band fully localizes. The outer, nontrivial bands leave behind extended states which levitate towards each other and annihilate around $W\approx 8$. 
This conventional behavior parallels that observed in earlier works on disordered topological phases \cite{metal_disorder_1, metal_disorder_2}.
Also for $v \gtrsim 5.65$, the system's behavior is typical: all bands are trivial, they all localize in the presence of disorder, and no levitation and annihilation can be seen.

A qualitatively different behavior, however, can be seen for intermediate coupling, $3.3 \lesssim v \lesssim 5.65$.
Here, instead of localizing, the middle band produces two sets of extended states, located symmetrically around $E=0$.
These proceed to levitate away from $E=0$, and annihilate with the extended states that originated from the top-most and bottom-most bands [see Fig.~\ref{fig:evolution}(b2)].
Contrary to the expectation that critical states with opposite topological charges annihilate first with their nearest neighbors in energy, the annihilation process in the present model is more intricate, suggesting sensitivity to the type of disorder.

The unconventional extended states emerging from the middle band carry opposite Chern numbers, as we have checked by repeating the calculation with open boundary conditions, and also by computing the scattering-matrix topological invariant~\cite{Brouwer,Braeunlich2010,Cosma_2012}. 
Further, by performing three-terminal transport simulations and a finite-size scaling analysis (shown in Appendix~\ref{sec:transport_simulations}), we have checked that these nontrivial extended states appear to persist for any disorder strength $W\neq 0$, no matter how small. 
This, however, does not rule out the possibility that in the thermodynamic limit these extended states only appear at some small but nonzero disorder only, as in the case of the topological Anderson insulator~\cite{tai_numerics,tai_theory}. To rule this out we use appropriate topological invariants below. 

\emph{Topological fine structure} --- 
We now show that the phenomenon of topological extended states emerging from the trivial band, observed in our example at coupling $v\approx 4.5$, persists also in the thermodynamic limit. 
For this, we employ a recently introduced tool for computing real-space topological invariants, the ``spectral localizer'', and quantities computed from it, the localizer index, and the localizer gap \cite{spectral_localizer, LORING2015383, Loring_finite_volume, Loring_Hermann_2, Cosma_localizer, Schulz-Baldes_2021, Liu_prb, Schulz-Baldes_2022, Selma_localizer}.
We rely on revealing what we call the topological fine structure of the middle band, captured by these quantities. We briefly summarize these concepts below, and clarify how they can show robustly delocalized states inside a seemingly topologically trivial (Chern number 0) band. We then calculate these quantities for our model, and prove that for $v\approx 4.5$, there must exist at least two energies in the middle band where eigenstates cannot become localized by weak disorder.

Our starting point is the spectral localizer ${\cal L} (r, E)$
\cite{spectral_localizer, LORING2015383, Loring_finite_volume, Loring_Hermann_2, Cosma_localizer, Schulz-Baldes_2021, Liu_prb, Schulz-Baldes_2022, Selma_localizer}, 
a matrix-valued function, defined for a finite-size sample of the 2D system, class A, with open boundary conditions. 
It is a continuous function of a reference position $\mathbf{r} = (r_x, r_y)$, which
is encoded via a complex number $r=r_x + i r_y$,  and of the energy $E$.
It is a Hermitian matrix of size $2Nm \times 2Nm$ for a system with $N$ unit cells and $m$ orbitals,   
\begin{equation}
    {\cal L}(r,E)=\begin{pmatrix}
        H-E&\kappa(X-iY-r^{*})\\
        \kappa(X+iY-r)&-H+E
    \end{pmatrix}.
    \label{eq:spectral_localizer}
\end{equation}
Here, $H$ is the Hamiltonian matrix of the finite-size system, $X$ and $Y$ are the matrices of the position operators. The dimensional parameter $\kappa$ has to be chosen to be small enough, see the Appendix ~\ref{sec:spectral gap scaling} and Ref.~\cite{spectral_localizer} for a more detailed description. In our case and with our units this was fulfilled by setting $\kappa=0.25$. 

The first piece of information provided by the spectral localizer matrix is the so-called localizer index $Q(r, E)$.  
It is defined as the matrix signature  (${\rm sig}$, the number of positive eigenvalues minus the number of negative eigenvalues) of the spectral localizer \cite{LORING2015383}, 
\begin{align}
    Q(r, E) &= \frac{1}{2}{\rm sig}[\mathcal{L}(r, E)].
    \label{eq:localizer_index}
\end{align}

At energy $E$ in a spectral gap or mobility gap, the localizer index is identical to the number of chiral edge modes at the system boundary for any
value of $\mathbf{r}$ chosen deep within the system's bulk \cite{spectral_localizer, LORING2015383, Loring_Hermann_2, Loring_finite_volume}. 
Thus, under these conditions, it contains the same information as the
Chern number, giving the value of the Hall conductance \cite{10.1063/1.5094300}, 
and it does not depend on the reference position $\mathbf{r}$, as long as $\mathbf{r}$ is chosen deep in the bulk.

We want to use the localizer index $Q(r, E)$ for energies $E$ inside an energy band; here, its value depends on the reference position $\mathbf{r}$ inside the bulk, but only if there are reference positions  where at least one of the eigenvalues of $\mathcal{L}(r, E)$ is 0. 
It can happen that there are no such values of $\mathbf{r}$, and the localizer index is constant throughout the bulk; under what circumstances this is expected to occur is currently not understood \cite{spectral_localizer, LORING2015383, Loring_Hermann_2, Loring_finite_volume}.

To check whether the localizer index is independent of the reference position, we use the localizer gap $g_{\cal L}$~\cite{spectral_localizer, LORING2015383, Loring_Hermann_2, Loring_finite_volume}. This is at a given energy $E$ the smallest absolute value of any of ${\cal L}(r, E)$'s eigenvalues, i.e. the shortest distance from an eigenvalue of ${\cal L}(r, E)$ to zero, at any $\mathbf{r}$ inside the bulk, i.e., 
\begin{align}
    g_{\cal L}(E) &= \!\! \min_{\mathbf{r} \in \text{bulk}} g_{\cal L}(r, E);& 
    g_{\cal L}(r, E) &= \!\!\! \min_{\lambda \in \sigma[\mathcal{L}(r, E)]} \abs{\lambda},
    \label{eq:localizer_gap}
\end{align}
where $\sigma[\mathcal{L}]$ is the set of eigenvalues of $\mathcal{L}$. 
We characterize the topological fine structure of a band using the localizer index $Q(r, E)$, evaluated at energies where the localizer gap $g_\mathcal{L}(E)$ is nonzero, and hence, the index is independent of $r$, 
\begin{align}
    Q(E) &= Q(r, E) \quad \text{when } g_\mathcal{L}(E) > 0. 
\end{align}

Besides ensuring the localizer index depends only on energy and not reference position, the localizer gap $g_\mathcal{L} (E)$ also quantifies the robustness of the index. 
Given Eq.~\eqref{eq:localizer_index}, the index $Q(r, E)$ cannot change under any perturbation, unless that perturbation is large enough to close the localizer gap, meaning that $g_{\cal L}(E)$ is reduced to zero. This is simply a consequence of Weyl's inequality: when perturbing $H\to \widetilde{H}$ and thus ${\cal L} \to \widetilde{\cal L}$, $Q$ is unchanged as long as $|| \widetilde{\cal L} - {\cal L} || < g_{\cal L}$. Here, $||\cdot ||$ is the 2-norm of a matrix, the modulus of its largest eigenvalue.

The localizer index $Q(E)$ can reveal topological fine structure of an energy band of a clean 2D system. It predicts the number of chiral edge states that would be seen at $E$ on a large enough sample with open boundary conditions, if weak disorder was added so as to localize the bulk states at energy $E$. This follows from the previous paragraph, with the weak disorder being treated as a perturbation.

If an energy band includes energy values $E_1 < E_2$ where the localizer indices differ, $Q(E_1) \neq Q(E_2)$, then there must exist an energy value between them, $E_1< E < E_2$, where energy eigenstates remain extended under weak disorder. This is necessarily true, because in the weakly disordered system the only way the number of chiral edge states at $E_1$ and $E_2$ can be different is if there is an energy between these values where the bulk gap closes (a topological phase transition in the spectrum of the disordered system). 
This shows how robustly and nontrivial extended states can emerge from a seemingly topologically trivial band (Chern number 0). 

\begin{figure}
\centering
\includegraphics[width=\linewidth]{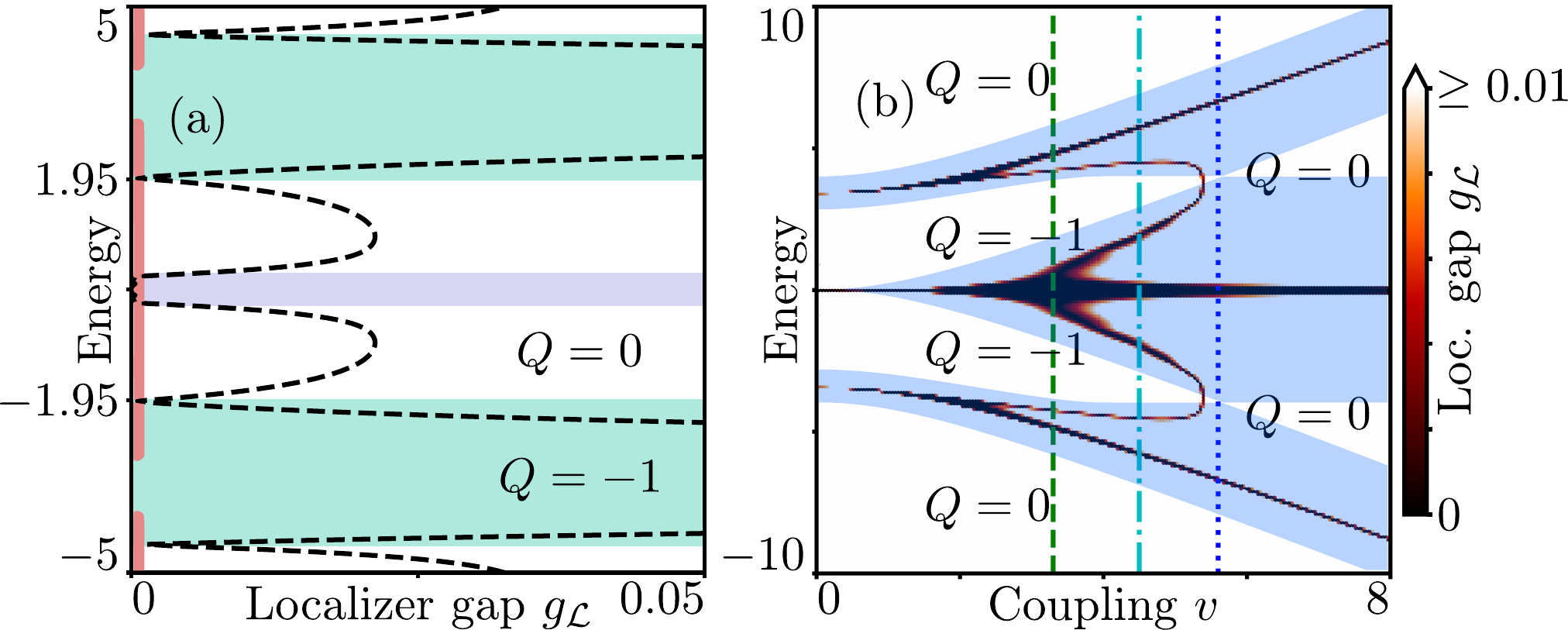}
\caption{
(a) Topological fine structure of the central energy band at coupling $v=4.5$: Localizer gap $g_\mathcal{L}$ (dashed line, horizontal axis) and localizer index $Q$ (shading and labels).  
Horizontal gray regions: energy ranges in which the $g_{\cal L}=0$, thus $Q(E)$ cannot be evaluated.
Vertical red bars on left axis: energy intervals of the bulk bands.
(b) Phase diagram of the topological fine structure revealed using the localizer gap $g_{\cal L}$ (shading in black to yellow/white), obtained from a $10\times 10$ unit cell system, plotted versus energy $E$ and coupling $v$.
Since a change of $Q$ requires a closing of the localizer gap, regions delimited by black lines have uniform $Q$ values, as shown.
Semi-transparent light blue shaded areas: energy intervals of the bands. 
Vertical dashed line: the critical point $v\approx 3.3$ at which the topological phase transition of the middle band occurs with no closing of the bulk gaps -- as obtained from the transport simulations. 
Dash-dotted line: $v=4.5$, corresponding to panel a). 
Dotted line: topological phase transition with a change in the Chern numbers, and closing of the bulk band gaps.
All plots are generated from the clean system with $W=0$ and $\kappa=0.25$, and $g_{\cal L}$ was computed taking a grid of $40\times 40$ reference positions $\mathbf{r}$ in the Wigner-Seitz cell.
\label{fig:localizer_gap}
}
\end{figure}

\emph{Topological fine structure in our example} --- 
We now apply the formalism above to our model, the Hamiltonian of Eq.~\eqref{eq:Hamiltonian}, to show that the delocalized states we observe in Fig.~\ref{fig:evolution} are delocalized at weak disorder in the thermodynamic limit. 
For the numerical work, 
we took square shaped systems of $20\times 20$ unit cells with open boundary conditions (we verified that repeating the calculation for larger system sizes does not significantly alter the result). 
We chose numerous energy values $-10 < E < 10$, and 
calculated the spectral localizer, Eq.~\eqref{eq:spectral_localizer}, and its eigenvalues of smallest magnitude over a
grid of 40$\times$40 reference positions $\mathbf{r}$ in the central Wigner-Seitz unit cell. 
Thus we obtained the localizer gap, Eq.~\eqref{eq:localizer_gap}. 
We note that, as proven in 
Ref.~\cite{Loring_finite_volume}
and verified by us in the Appendix~\ref{sec:spectral gap scaling}, for a fixed set of Hamiltonian parameters and $\mathbf{r}$ in the bulk, $g_{\cal L}(r, E)$ converges as system size is increased, thus only the behavior of the system close to the position encoded in $r$ influences $g_{\cal L}(r, E)$.
We picked some $v, E$ values where we also fully diagonalized the spectral localizer to obtain the localizer index $Q(E)$. Since this topological invariant cannot change as long as the localizer gap is nonzero, this numerically costly procedure was only needed at a few $(v, E)$ pairs.   

First, we show our main result, the topological delocalized states at $v=4.5$,  in Fig.~\ref{fig:localizer_gap}(a). 
Here, in both spectral gaps we find $Q=-1$, consistent with the Chern numbers of the bands.
Remarkably, the localizer gap remains open (e.g. $g_{\cal L} > 0.02$) even for energies \emph{within} the central band, meaning that the localizer index $Q(E)$ remains well defined for most energies in the band. 
Moreover, the index changes from $Q=-1$ to $Q=0$ and then back to $Q=-1$ as energy is scanned through the middle band.
Thus, there are two topological phase transitions in that band: this is the topological fine structure, predicting two energies where robustly extended states must occur. 
Here, these transitions are at $E \simeq \pm 1.95$, which agrees with Fig~\ref{fig:evolution}(b2). 
We note that the precise energy location of these transitions can, in principle, be weakly affected by the choice of orbital embedding (for instance, small shifts of sublattice orbital positions within the unit cell). However, in practice this dependence only leads to minor quantitative shifts and cannot alter the overall structure of the topological fine structure.

Second, we show a full phase diagram of the topological fine structure of Hamiltonian, Eq.~\eqref{eq:Hamiltonian}, obtained by repeating the calculation above for many values of $v$, in Fig.~\ref{fig:localizer_gap}(b). 
We find the nontrivial topological fine structure of the middle band  throughout the interval $3.3 \lesssim v \lesssim 5.65$. This is consistent with our weak disorder numerics which show robustly extended states in the middle band also when adding additional perturbations to the Hamiltonian, such as shifting the energy of the band (see Appendix~\ref{sec:mu_2}). 
Note that the mismatch between the localizer gap closing and the closing of bulk band gaps is a technical issue of choices of $\kappa$, see Appendix~\ref{sec:spectral gap scaling} for more details. Note that the localizer gap is 0 close to $E=0$ for all values of $v$ we consider, however, there is no topological phase transition in the spectrum here, no robustly extended eigenstates are expected, since the localizer index does not change across $E=0$.   

We emphasize that the emergence of extended states from the trivial band is the consequence of a topological phase transition occurring as $v$ crosses a critical value, between 3.1 and 3.3, as seen in Fig.~\ref{fig:localizer_gap}(b) as well as the additional transmission calculation in Appendix~\ref{sec:transport_simulations}. However, this transition does \emph{not} involve a closing of the bulk gaps of the system, and it is not associated to a change in the Chern numbers of the bands.  

To place our approach in a broader context, we briefly compare it to other methods used to identify the energies of topological critical states. 
In disordered systems, methods are based on finite-size scaling of transport or multifractal analysis, which can provide quantitatively accurate results, albeit at an increased computational cost. 
This motivates the use of clean-system approaches. A commonly used strategy is based on the intrinsic anomalous Hall conductivity, where critical energies are estimated from the condition $\sigma_{xy}^{\text{Int}}=(2i+1)/2$ (with $\sigma_{xy}^{\text{Int}}$ being the integral of Berry curvature over the energy interval and $i$ representing the $i$th critical state). However, changing orbital position within the unit cell will provide different critical energies, thus making energy intervals. 
The spectral localizer approach adopted here offers a complementary perspective by detecting topological charges through changes in a local index in the clean limit, thereby directly identifying the energy region where the topological fine structure emerges, without requiring significantly large system sizes and many disorder realizations. While it also depends on the choice of position operators, in practice it yields a well-defined ( but narrow) energy window for critical states. 

We note that the localizer gap at $E=0$ vanishes for all $v$, indicating the existence of a zero mode of the spectral localizer, $\mathcal{L}|\psi_\mathcal{L}\rangle=0$. For $v=0$, this can be understood from the decoupled trivial flat band, whose perfectly localized zero-energy eigenstates satisfy both $H|\psi\rangle=0$ and $(X+iY-r)|\psi\rangle=0$, thereby yielding an exact zero mode of the localizer. 
Upon increasing $v$, the central band acquires a nonzero bandwidth, and its eigenstates become extended Bloch waves, so this simple mechanism no longer applies. 
The persistence of the vanishing localizer gap at finite $v$ therefore suggests that the zero mode is not solely tied to the perfectly localized flat-band states. We believe this behavior is non-generic and specific to our Hamiltonian rather than a universal feature.

\emph{Conclusion and outlook} --- 
We have shown that a seemingly topologically trivial band can exhibit a nontrivial topological fine structure, with a generalization of the Chern number taking different values inside the band. 
Such a fine structure implies that the band hosts eigenstates that are robustly extended under weak disorder. 
The generalization of the Chern number capturing the topological fine structure is the localizer index and the localizer gap of the spectral localizer. 

Our work will motivate future research in several directions. 
Saliently, while we have only considered class A in two dimensions, it is an intriguing open question what topological fine structure may exist in other symmetry classes and dimensions. In this context we note that topological markers have recently been generalized to odd-dimensional systems \cite{PhysRevLett.129.277601}. 

Another direction worth exploring is the interplay between the fine structure topology and interactions. This, we conjecture, may lead to fractional Chern insulators \cite{fci, Liu2023}, which usually, but not always \cite{FCI_zero_berry}, require a band with nonzero Chern number. In the present context we envision that stable states may occur at unconventional filling fractions. In particular, Laughlin-like states may instead occur at even denominator band filling (corresponding to odd denominator filling of the effective fine structure band).

On a more practical level, it would be interesting to identify further models exhibiting the topological fine structure phenomenology. Given the generality of our argumentation such models should indeed be ubiquitous. In particular, it would be interesting to find a minimal model, presumably featuring only two energy bands. Nevertheless, we emphasize that the present three band setting is also realistic. In fact, attaching a trivial flat band to a Chern insulator has already been realized in photonic systems~\cite{experiment1, experiment3, experiment2}. We expect that our work will motivate further experimental research in this direction.

Notions of band structure topology have profoundly changed the way we understand phases of matter and altered the paradigm of localization. Topological fine structure provides a natural next level of understanding of these fundamental concepts. 

\emph{Acknowledgements} ---
We thank Ulrike Nitzsche for technical assistance. 
ICF acknowledges support from the Deutsche Forschungsgemeinschaft (DFG, German Research Foundation) under Germany's Excellence Strategy through the W\"{u}rzburg-Dresden Cluster of Excellence on Complexity and Topology in Quantum Matter -- \emph{ct.qmat} (EXC 2147, project-ids 390858490 and 392019). 
HL and EJB were supported by the Swedish Research Council (VR, grant 2018-00313), the Wallenberg Academy Fellows program (2018.0460) of the Knut and Alice Wallenberg Foundation, and the G\"oran Gustafsson Foundation for Research in Natural Sciences and Medicine.
JKA acknowledges support by the National Research Development and Innovation Office (NKFIH) through the OTKA Grant FK 132146. 
%Janos2026
This work was supported by the HUN-REN Hungarian Research Network through the Supported Research Groups Programme, HUN-REN-BME-BCE Quantum Technology Research Group (TKCS-2024/34).

\appendix

\section{Real space Hamiltonian}
\label{sec:real space lattice}
\begin{figure}
\centering
\includegraphics[width=0.5\linewidth]{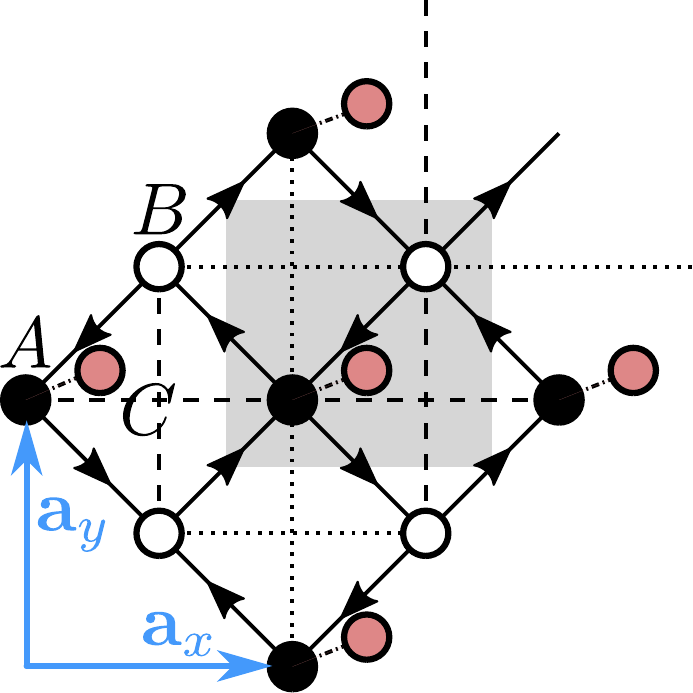}
\caption{Sketch of the model in real space. Here $A$ and $B$ represent the sublattices of the two band topological model, respectively. $C$ indicate the extra site coupled to sublattice $A$. 
\label{fig:system}
}
\end{figure}

In Fig.~\ref{fig:system}, we provide the information of the lattice.
In real space, the Hamiltonian reads
\begin{eqnarray}
    H&&=(1+i)\sum_{\mathbf{r}}(|A_{\mathbf{r}}\rangle \langle B_{\mathbf{r}}|+|A_{\mathbf{r}}\rangle \langle B_{\mathbf{r}+\mathbf{a}_x}|+|A_{\mathbf{r}}\rangle \langle B_{\mathbf{r}+\mathbf{a}_y}|\nonumber\\
    &&+i |A_{\mathbf{r}}\rangle \langle B_{\mathbf{r}+\mathbf{a}_x+\mathbf{a}_y}|)+\sum_{\mathbf{r}}(|A_{\mathbf{r}}\rangle \langle A_{\mathbf{r}+\mathbf{a}_x}|-|A_{\mathbf{r}}\rangle \langle A_{\mathbf{r}+\mathbf{a}_y}|\nonumber\\
    && -|B_{\mathbf{r}}\rangle \langle B_{\mathbf{r}+\mathbf{a}_x}|+|B_{\mathbf{r}}\rangle \langle B_{\mathbf{r}+\mathbf{a}_y}|)+v\sum_{\mathbf{r}}|A_{\mathbf{r}}\rangle \langle C_{\mathbf{r}}|+H.c.,\nonumber\\
\end{eqnarray}
with $A$ and $B$ denoting orbitals corresponding to the Chern insulator, $C$ to the trivial insulator, and $\mathbf{r} = n_x \mathbf{a}_x +  n_y \mathbf{a}_y$ is position on the two-dimensional square lattice, with $n_x, n_y \in \mathbb{Z}$. Our unit of length is the lattice constant: $\abs{\mathbf{a}_x} = \abs{\mathbf{a}_y}=1$. 

\section{Details on transport setting}
\label{sec:transport setting}
\begin{figure}
\centering
\includegraphics[width=\linewidth]{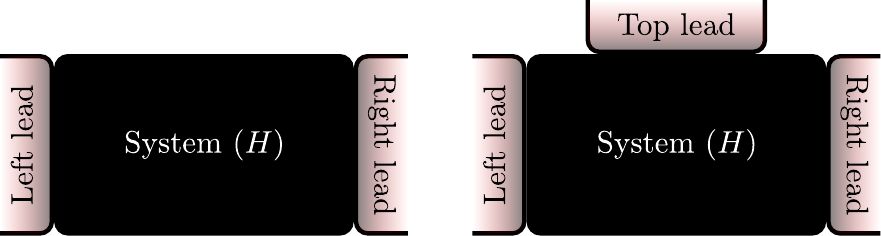}
\caption{The left and the right panel are the two-terminal and three-terminal transport setting, respectively. 
Here, all leads are ideal (with ballistic transports) and semi-infinite long.
\label{fig:transport setting}
}
\end{figure}

Here, we give more details on the transport simulations we used. 
For the two-terminal transport setting, as shown in the left panel of Fig.~\ref{fig:transport setting}, the scattering matrix is~\cite{Carlo_review}
\begin{equation}
    \begin{pmatrix}
        \Psi_{L,o}\\
        \Psi_{R,o}
    \end{pmatrix}=
    \begin{pmatrix}
    r&t'\\t&r'    
    \end{pmatrix}
    \begin{pmatrix}
        \Psi_{L,i}\\
        \Psi_{R,i}
    \end{pmatrix}=S_{2-\text{terminal}}    
    \begin{pmatrix}
        \Psi_{L,i}\\
        \Psi_{R,i}
    \end{pmatrix},
\end{equation}
with $\Psi_{L, i}$ ($\Psi_{R, i}$) being the input modes at left (right) boundary and $\Psi_{L, o}$ ($\Psi_{R, o}$) being the outgoing modes at left (right) boundary. Then by definition $r$ ($r'$) and $t'$ (t) are the reflection and transmission matrix to the left (right) boundary, respectively.
Then, the two-terminal transmission is $G=\text{tr}[tt^\dagger]=\text{tr}[t't'^\dagger]$. 

Attaching a third lead at the top boundary gives the three-terminal setting (see the right panel in Fig.~\ref{fig:transport setting}). 
Then the scattering matrix is 
\begin{equation}
S_{3-\text{terminal}}=\begin{pmatrix}
r_{LL}&t_{LT}&t_{LR}\\
t_{TL}&r_{TT}&t_{TR}\\
t_{RL}&t_{RT}&r_{RR}
\end{pmatrix}.  
\end{equation}
Here, $L$, $T$, and $R$ represent the left, the top, and the right lead, respectively. 
Then, we have $G_{L\rightarrow R}=\text{tr}[t_{RL}t_{RL}^\dagger]$ and $G_{R\rightarrow L}=\text{tr}[t_{LR}t_{LR}^\dagger]$.
Since the top lead absorbs the transmission between the left and right lead along the top boundary (i.e., the chiral edge state), $G_{L\rightarrow R}\neq G_{R\rightarrow L}$ in general.

\section{Additional transport simulations}
\label{sec:transport_simulations}

\begin{figure}
\centering
\includegraphics[width=\linewidth]{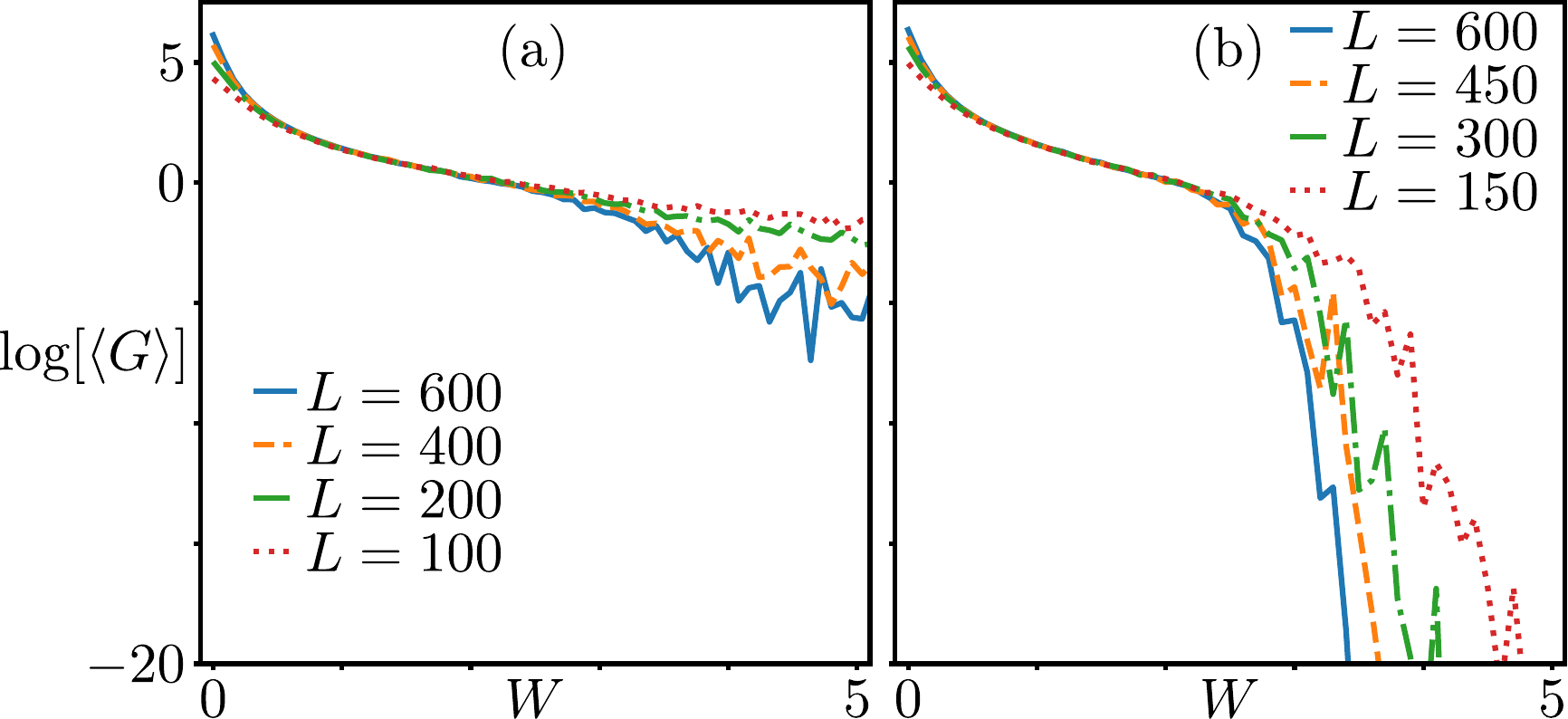}
\caption{The two-terminal transmission $G$ as a function of disorder for energy $E=0$ at $v=4$ (a) and $v=8$ (b) with different system sizes, respectively. 
Here, all plots are with an aspect ratio $1$ and $50$ disorder realizations.
\label{fig:two_terminal_conductance}
}
\end{figure}

We first present a finite size scaling analysis of the two-terminal transmission for $E=0$ at $v=4$ (where the trivial band splits) and $v=8$ (where the trivial band is localized by disorder).
In Fig.~\ref{fig:two_terminal_conductance}(a) and (b), it shows the transmission $G$ monotonically decreases with the disorder strength $W$, suggesting that no localization-delocalization transition takes place. 
This is consistent with the absence of a topological Anderson insulator behavior, in which topologically extended states present at $E=0$ would first localize, and then a topological phase transition would occur afterwards, at a slightly higher disorder strength. 

\begin{figure}
\centering
\includegraphics[width=0.5\linewidth]{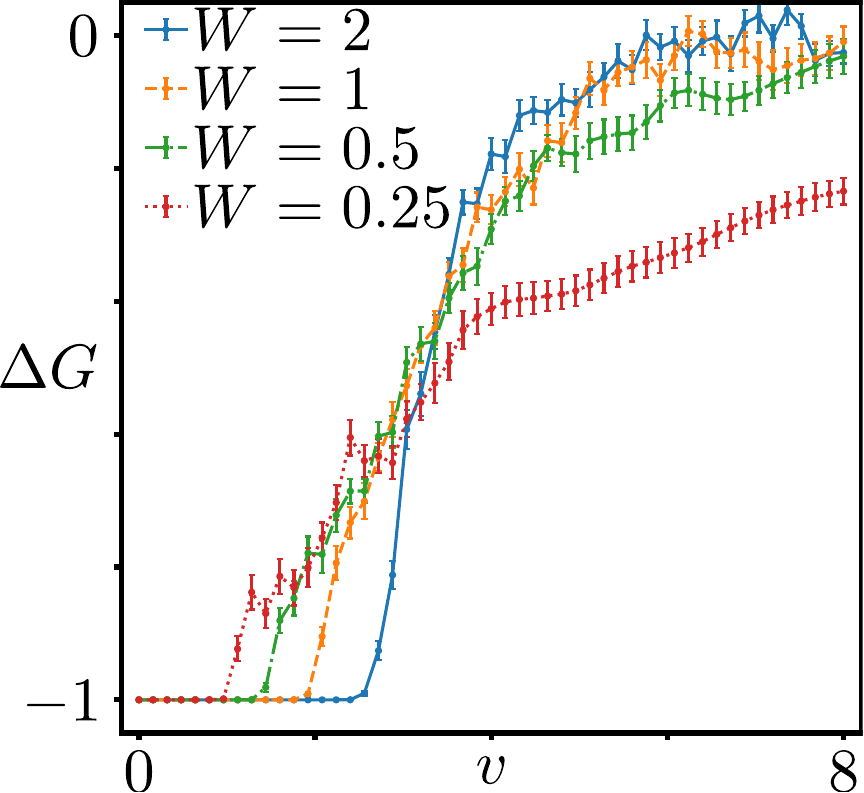}
\caption{The difference of the three-terminal transmission $\Delta G$ at $E = 0$ as a
function of $v$ at different disorder strength with $L = 400$.
Here, all plots are with an aspect ratio $1$ and $50$ disorder realizations.
\label{fig:three_terminal_conductance}
}
\end{figure}

\begin{figure}
\centering
\includegraphics[width=\linewidth]{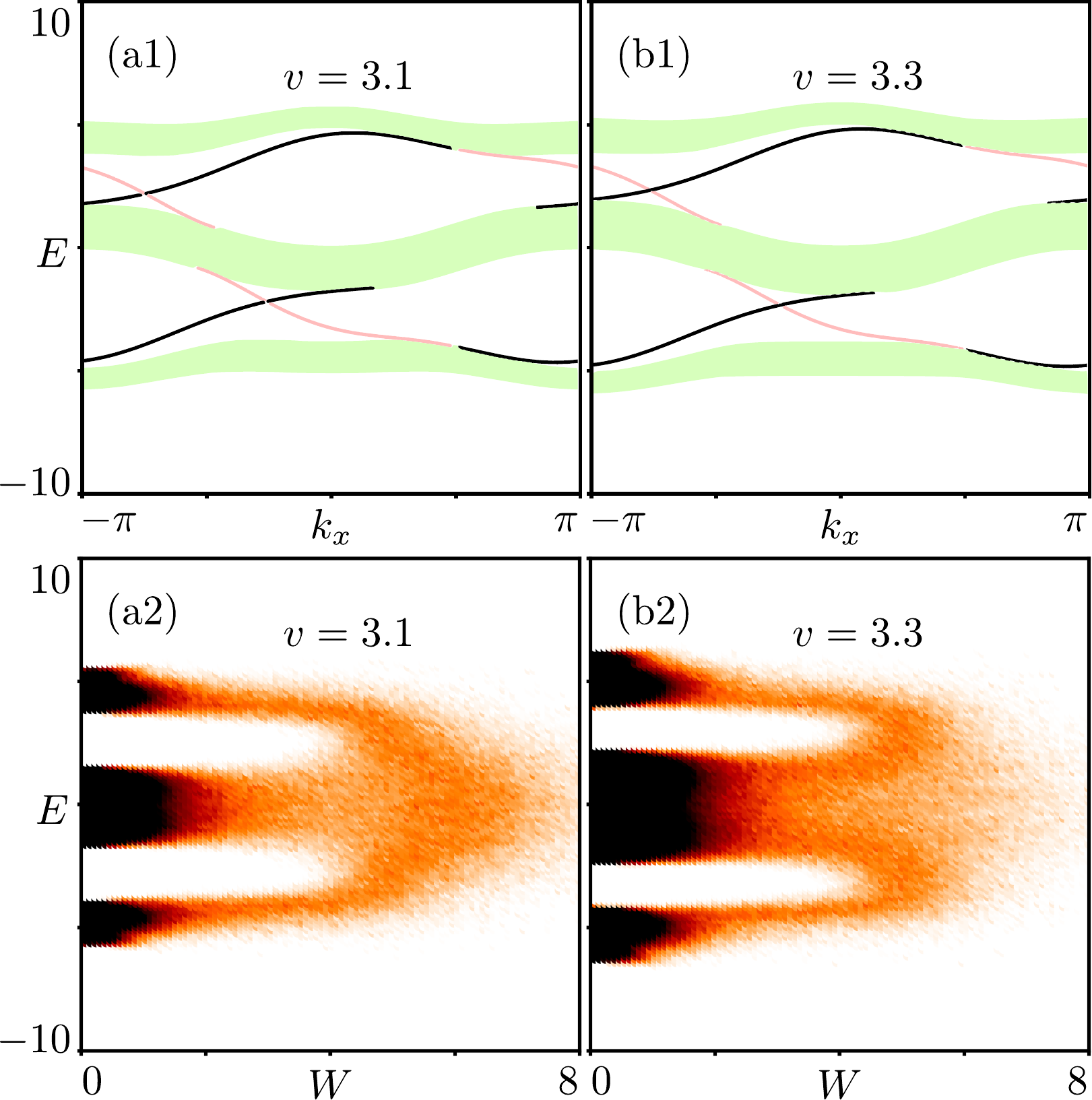}
\caption{Top panels: bandstructure of the model in the absence of disorder. We use a ribbon geometry, infinite along the $x$-direction and consisting of 60 unit cells along $y$. Black, red, and green colors indicate states localized at the top edge, bottom edge, and in the bulk, respectively.
Bottom panels:
Two-terminal transmission probability, plotted as a function of energy $E$ and disorder strength $W$, for a $60\times60$ unit cell system with periodic boundary conditions in the $y$-direction.
Darker regions correspond to the presence of extended bulk states, helping us track the levitation and annihilation process. Each point is obtained by averaging over 50 independent disorder realizations.
\label{fig:transmission_critical}
}
\end{figure}

The three-terminal transport setting can be used to identify the presence or absence of chiral edge states at weak disorder, even though in this situation chiral edge states might hybridize with the conducting bulk. 
Here, we expect the transmission within the bulk to be roughly the same both in the left-to-right and in the right-to-left directions. 
Then the only difference comes from the bottom edge, where the presence of chiral edge states leads $\Delta G = G_{R\rightarrow L}-G_{L\rightarrow R}$ to be nonzero, depending on the presence of edge states. 
Thus, if changing $v$ results in a topological phase transition within the trivial band, $|\Delta G|$ at a energy between these two subbands should jump from $1$ to $0$ at arbitrary weak disorder in the thermodynamic limit.
From a realistic perspective, $W\rightarrow 0$ will lead to strong finite size effects, since the localization length becomes much larger than the system size.
But as shown in Fig.~\ref{fig:three_terminal_conductance}, there is still a jump of $\Delta G$ from $-1$ to roughly $0$ for $W\in [0.5, 2]$, which happens around $v\approx 3.2$. 
This has a good agreement with the two-terminal transport simulation  (see Fig~\ref{fig:transmission_critical}), where the separation of the trivial band occurs in between $v=3.1$ and $3.3$.

\section{Winding number calculations at strong disorder}
\label{sec:winding number}

\begin{figure}
\centering
\includegraphics[width=0.5\linewidth]{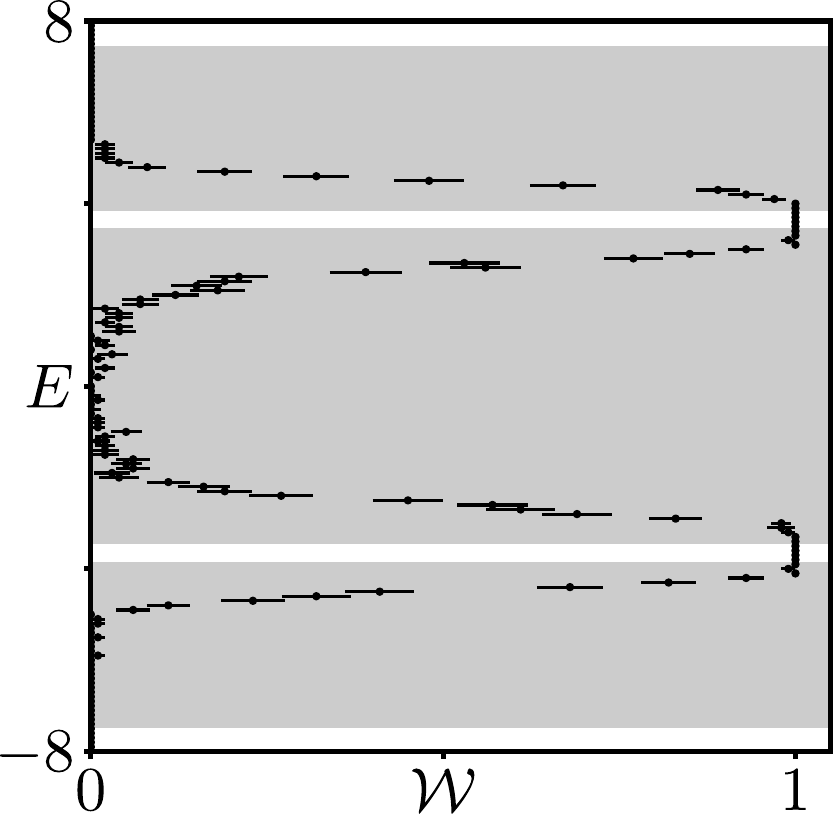}
\caption{
Winding number $\mathcal{W}$ as a function of energy $E$. Here, the gray and white region represent the bulk bands and the gaps of the clean system, respectively. This plot is with  $v=5$, $W=3.5$, a system size $50\times 50$ unit cells, and $100$ disorder realizations.
\label{fig:winding_number}
}
\end{figure}

In this section, we use winding number of the reflection matrix to evaluate the topological property of the trivial band at strong disorder strength~\cite{Brouwer, Braeunlich2010, Cosma_2012}, 
\begin{equation}
    \mathcal{W}=\frac{1}{2\pi}\int_{0}^{2\pi} \frac{d}{d\phi}\text{arg}[\text{det}[r(\phi)]]d\phi.\label{eq:winding_number}
\end{equation}
Here, $r$ is the reflection matrix of the system in a two terminal transport setting (see Fig.~\ref{fig:transport setting}), with periodic boundary conditions along the $y$-direction, and $\phi$ is the inserted magnetic flux, which changes from $0$ to $2\pi$. 
In this context, the bulk system has to be insulating to guarantee the unitarity of the reflection matrix. Otherwise, the winding number is ill-defined.

As shown in Fig.~\ref{fig:winding_number}, when the bulk is localized by strong disorder, there are two nonzero quantized plateaus ($E\in [3, 4.1]$ and $[-4.1, -3]$) with Winding number $\mathcal{W}=1$ contributed by the edge states. 
In other energy internals, $\mathcal{W}=0$ indicates the absence of edge states. 
Jumping from between $\mathcal{W}=1$ and $0$ indicates a mobility gap closure, where the presence of extended states ill-defines the winding number, which leads to a large errorbar in numerics. 

In this context, an alternative definition of the Chern number of an energy set (when disordered) is the difference between the winding number above and below the energy set (here, we mean the energy set of extended states). 
Then, the four extended states along the energy axis from bottom to top carry Chern number $-1$, $1$, $-1$, and $1$, respectively. 
We thus confirm all extended states are topologically nontrivial. 

\section{Localizer gap for {$v=2$}}
\label{sec:v2_2}

\begin{figure}[tb]
\centering
\includegraphics[width=0.5\linewidth]{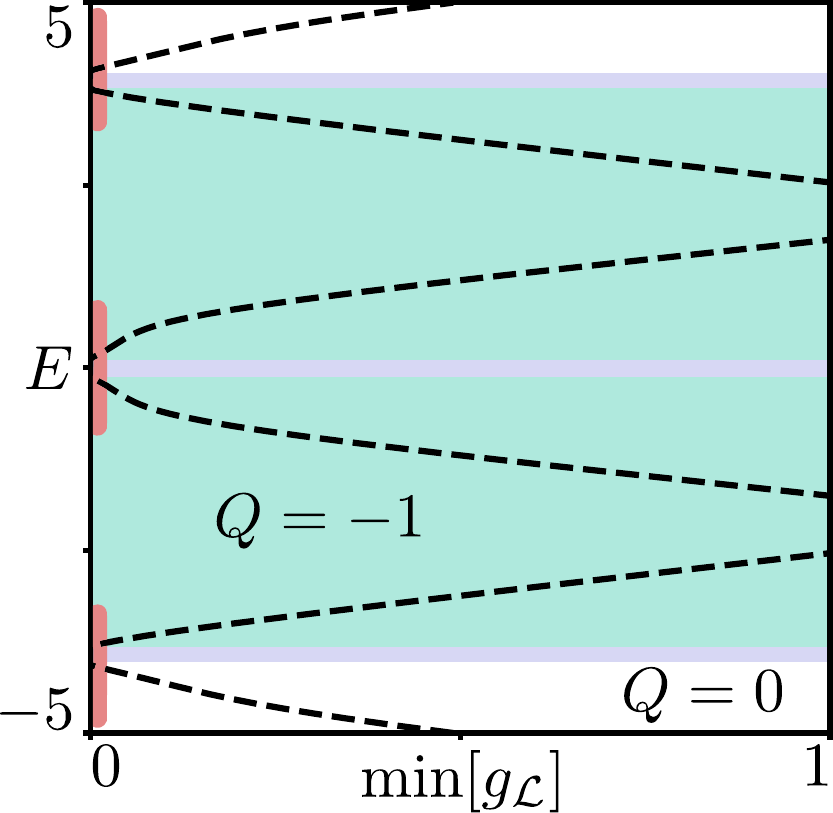}
\caption{
The smallest localizer gap as a function of energy for $v=2$. This plot is with a system size $20\times 20$ unit cells and $g_{\cal L}$ is over all positions in the middle Wigner-Seitz unit cell with a grid of $40\times 40$ points.
\label{fig:v2_2}
}
\end{figure}

In the main text, we have shown the localizer gap of $v=4.5$, where the change from $Q=-1$ to $Q=0$ within the middle band validates the existence of extended states originated from the trivial band.
Here, we explore the case of $v=2$. 
As shown in the main text, the levitation and pair annihilation process only involves the top and bottom band. 
The middle band with ${\cal C}=0$ is localized by weak disorder.
Consequently, $g_{\cal L}$ as a function of $E$ in Fig.~\ref{fig:v2_2} shows a consistent behavior, where within the middle band, $Q$ is either $-1$ or ill-defined ($g_{\cal L}=0$), indicating the absence of extended states when disordered.

\section{More details of the spectral localizer and the choice of $\kappa$}
\label{sec:spectral gap scaling}

In this section, we provide more details on spectral localizer calculations, including how to choose the value of $\kappa$. 

In topological phases of matter, distinguishing between topologically distinct phases relies on topological invariants defined in momentum space. These invariants identify if there exists an obstruction preventing band structure of a system from smoothly deforming into an atomic insulator, where all eigenstates are perfectly localized. In real space, this corresponds to measure if a Hamiltonian $H$ can be smoothly deformed to commute with the position operators, without closing a gap or breaking a certain symmetry. One tool employed for this purpose is the spectral localizer rooted in $K-$theory. It identifies if a Hamiltonian can or cannot be adiabatically changed to commute with position operator matrices, through the signature of the spectral localizer matrix and its associated gap~\cite{LORING2015383, spectral_localizer}. Here, the localizer gap is defined by solving the eigenvalue equation of the spectral localizer at a given energy $E$ and position $\mathbf{r}$, as detailed in the main text.  
This indicates if the system has a state at this specific energy with a large weight at this position. 
If not, the localizer possesses a gap. 
Together with the matrix signature, it can serve as an efficient tool to distinguish different topological phases.

\begin{figure}
\centering
\includegraphics[width=\linewidth]{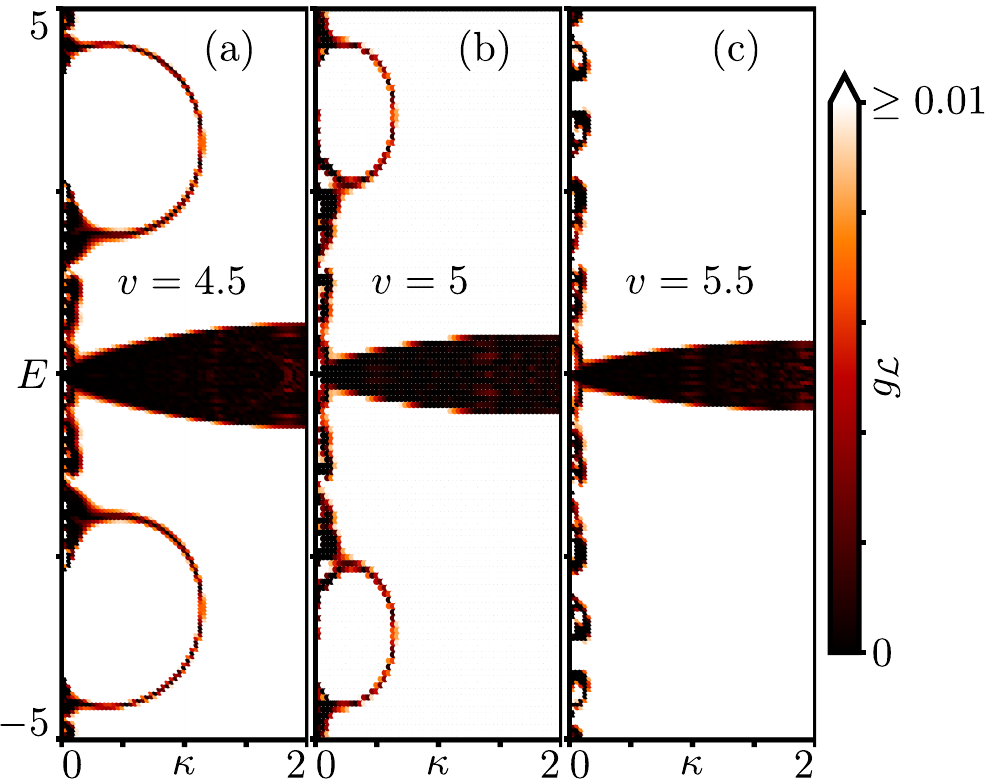}
\caption{The smallest localizer gap as a function of energy and $\kappa$ for different $v$. All plots are with a system size $10\times 10$ unit cells and $g_{\cal L}$ is over all positions in the middle Wigner-Seitz unit cell with a grid of $40\times 40$ points.
\label{fig:evolution_v_kappa}
}
\end{figure}

\begin{figure}
\centering
\includegraphics[width=0.5\linewidth]{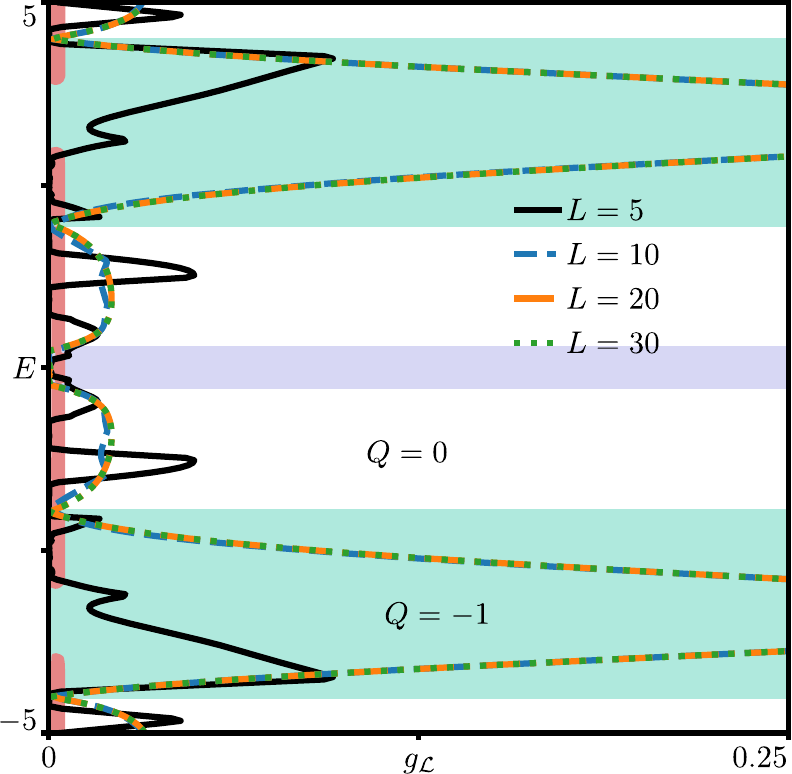}
\caption{Smallest value of the localizer gap $g_{\cal L}$ (dashed line) over all positions of the middle unit cells as a function of energy $E$ at $v=4.5$ with different system sizes. Here, we choose $\kappa=0.25$, a system size $L\times L$ unit cells, and a grid of $40\times 40$ points in the Wigner-Seitz unit cell.
\label{fig:spectral_gap_scaling}
}
\end{figure}

\begin{figure}
\centering
\includegraphics[width=0.5\linewidth]{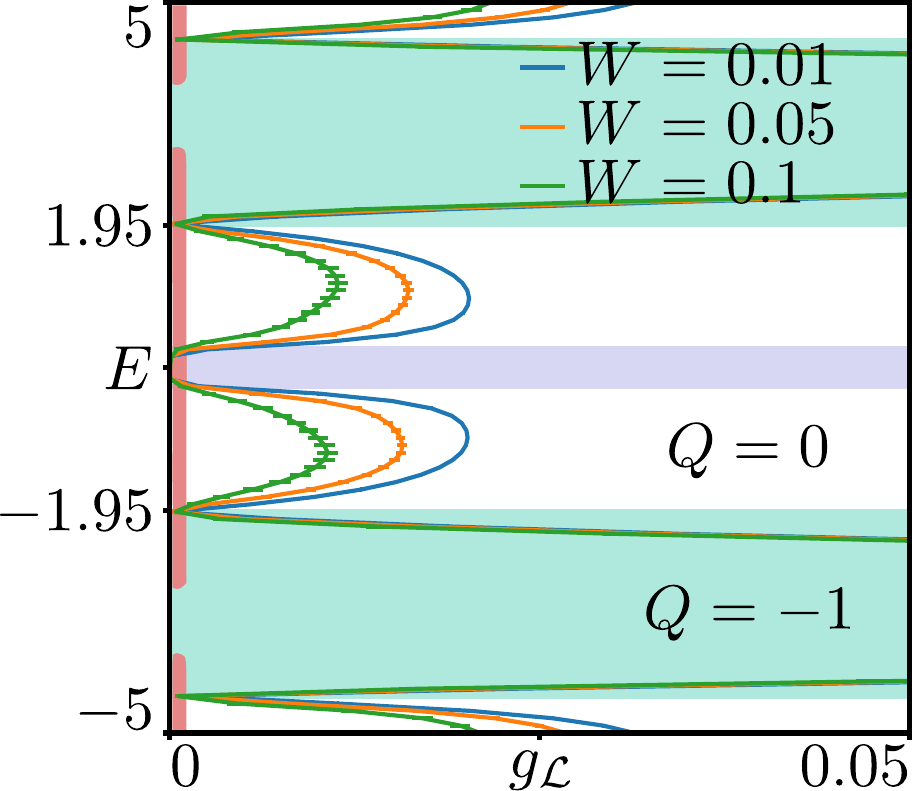}
\caption{The averaged localizer gap in the presence of disorder as a function of energy. 
Here we choose $v=4.5$, a system size $20\times 20$ unit cells, and 50 disorder realizations.
\label{fig:gl_disorder}
}
\end{figure}

For a finite system with $L\times W$ unit cells, the position operator $X$ and $Y$ are defined as diagonal matrices with matrix elements, in general, being real numbers. 
Here, for simplicity, we set all matrix elements to be integers varying from $0$ to $L-1$ and $W-1$, respectively. 

We then introduce the role $\kappa$ to finely modulate the balance between the position operator and the Hamiltonian matrix. When $\kappa$ is too small, the dominance of the Hamiltonian matrix over the localizer $\mathcal{L}$ is pronounced, resulting in a symmetric spectrum and a localizer index $Q=0$. 
Conversely, a large $\kappa$ downplays the influence of the Hamiltonian matrix and then the localizer only reflects the information of the position operator, which thus again yields a trivial localizer index.
In this sense, an optimal choice for $\kappa$ is to make $\kappa[H,X]$ and $\kappa[H,Y]$ sufficiently small. 
It positions in between the two contrasting extreme cases, which ensures that both the Hamiltonian matrix and the position operator contribute meaningfully to the overall localizer matrix
(An intuitive way for choosing a good $\kappa$ is to evaluate how the spectral localizer gap changes as a function of $\kappa$, as shown in Fig.~\ref{fig:evolution_v_kappa}.).
We have verified for this model, away from the bulk gap closing $v\approx 5.65$, $\kappa\in[0.25, 0.75]$ is good enough, and we expect it to hold in an even wider range.

Next, we show the localizer gap converges when increasing the system size. 
In Fig.~\ref{fig:spectral_gap_scaling}, by changing the system size from $5\times 5$ to $30\times 30$ unit cells, the smallest value of localizer gap over the middle unit cell becomes a fixed value. 
Thus, a finite system with $10\times 10$ unit cells is large enough to evaluate the topological fine structure. 

We now show that the localizer gap is robust to weak disorder. To be consistent with the result present in the main text, we choose $\nu=4.5$ and onsite disorder. The result is shown in Fig.~\ref{fig:gl_disorder}. It is clear that the localizer gap remains open with weak disorder, reflecting the robustness of the localizer invariant and the presence of the topological fine structure in the trivial band.

In the main text, we observe a discrepancy between the bulk gap closing and the localizer gap closing, which is a technical issue of choices of $\kappa$, as near the bulk gap closing ($v\approx 5.65$), we encounter difficulties in identifying a suitable $\kappa$. 
To show this, we systematically explore various values of $v$ and examine how the localizer gap changes with respect to $\kappa$. In Fig.~\ref{fig:evolution_v_kappa}(a-b), when deviating from the bulk gap closing, the presence of bubbles indicates an optimal $\kappa$ around $0.5$ for $v=4.5$ and $0.25$ for $v=5$. 
Further approaching $v\approx 5.65$, the bubble gradually diminishes, making it increasingly hard to find a suitable $\kappa$ for the system, as a signature of topological phase transitions.
After that, the clean system becomes fully trivial with all bands possessing zero Chern number and $Q=0$ for all energies. 
It is worth noting that close to $\kappa=0$, the dominance of the Hamiltonian matrix becomes pronounced, leading to the anticipation of inaccurate results as well.

\section{A case study for onsite potential $\mu=2$}
\label{sec:mu_2}

\begin{figure*}
\centering
\includegraphics[width=0.7\linewidth]{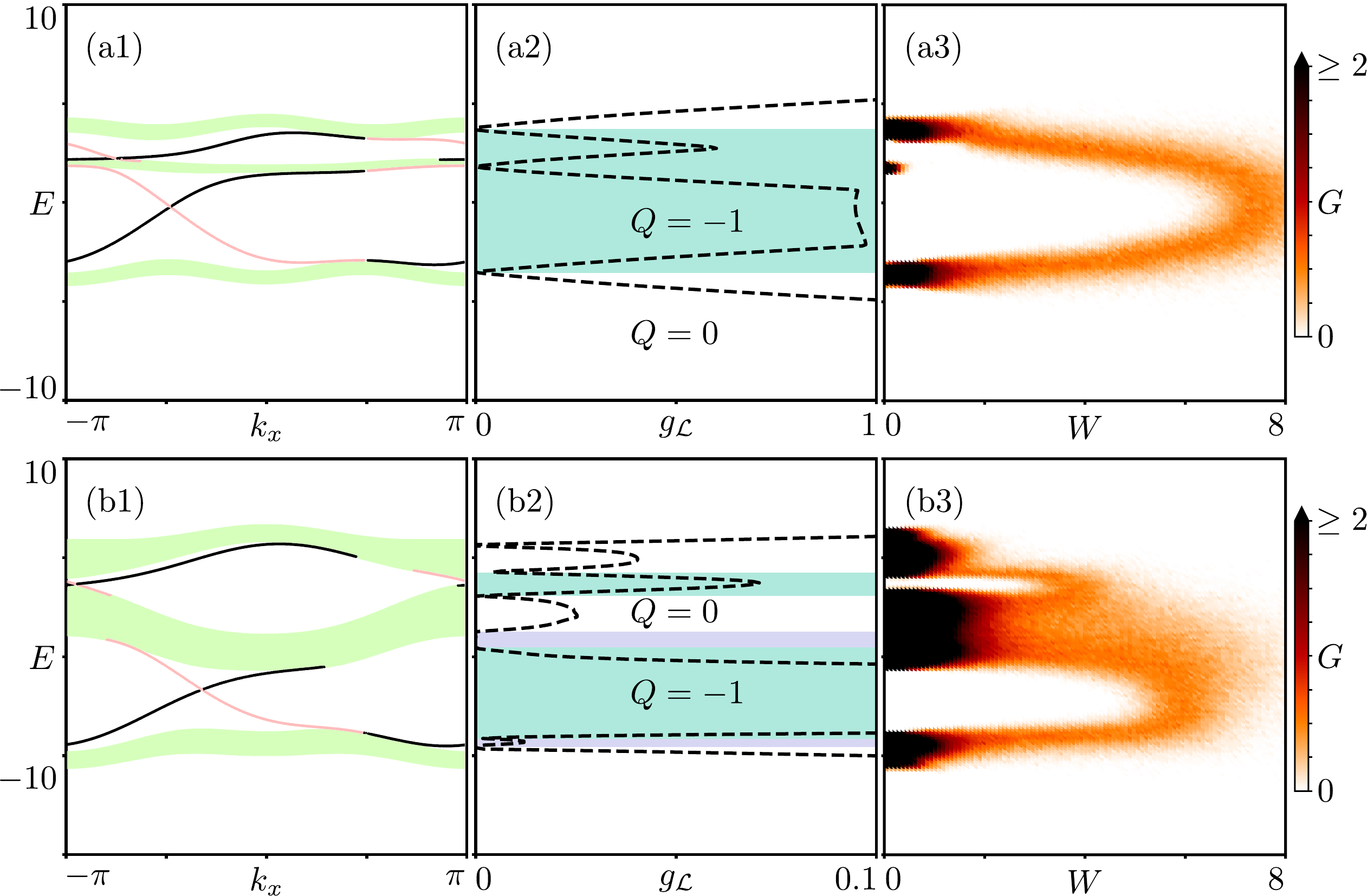}
\caption{top panels from (a1) to (a3) are the ribbon geometry spectrum, the smallest localizer gap over the middle unit cell as a function of energy, and the disorder phase diagram of transmission for $v=1$, respectively.
The bottom panels are their corresponding plots for $v=3.5$. 
The transmission plots are with a system size $60\times 60$ unit cells and $50$ disorder realizations. 
The invariant plots are with a system size $10\times 10$ unit cells, $\kappa=0.25$, and a grid of $40\times 40$ points in the middle Wigner-Seitz unit cell.
\label{fig:evolution_mu2}
}
\end{figure*}

In the main text, we have studied the system without considering on-site terms, where a chiral-mirror symmetry $\mathcal{U}H(k_x, k_y)\mathcal{U}^\dagger=-H(k_y, k_x)$ with $\mathcal{U}=\text{diag}[-1, 1, 1]$ is present. 
Here, we demonstrate the case without such a symmetry by setting the onsite potential $\mu=2$, which can be regarded as energy shift of the trivial flat band. 
As shown in Fig.~\ref{fig:evolution_mu2}(a1) and (b1), the system behaviors similar with the case of $\mu=0$, except the gap between the $2$nd and $3$rd band closes first. 
Before the gap closing, the top and bottom band have Chern number $+1$ and $-1$, respectively. And the middle band remains trivial. 

From localizer gap shown in Fig.~\ref{fig:evolution_mu2}(a2) and (b2), there are two different fine structures in the middle band. 
Then introducing disorder should lead to two different scenarios (see Fig.~\ref{fig:evolution_mu2}(a3) and (b3)). 
For $v=1$, the absence of changing in $Q$ within the middle band implies an ordinary levitation and annihilation taking place between the two band with Chern number $\pm1$ and the trivial band is self-localized. 
Conversely, for $v=3.5$, the localizer gap closing and reopening associated with $Q$ changing within the middle band indicates the presence of extended states.
In this case, disorder drives these two sets of extended states to meet and annihilate with top and bottom extended states, respectively.

\bibliography{reference}

@article{PhysRevLett.129.277601,
  title = {Local Topological Markers in Odd Spatial Dimensions and Their Application to Amorphous Topological Matter},
  author = {Hannukainen, Julia D. and Mart\'{\i}nez, Miguel F. and Bardarson, Jens H. and Kvorning, Thomas Klein},
  journal = {Phys. Rev. Lett.},
  volume = {129},
  issue = {27},
  pages = {277601},
  numpages = {6},
  year = {2022},
  month = {Dec},
  publisher = {American Physical Society},
  doi = {10.1103/PhysRevLett.129.277601},
  url = {https://link.aps.org/doi/10.1103/PhysRevLett.129.277601}
}

@Article{fqhe_zero_mag,
  author    = {Neupert, Titus and Santos, Luiz and Chamon, Claudio and Mudry, Christopher},
  journal   = {Phys. Rev. Lett.},
  title     = {Fractional Quantum {H}all States at Zero Magnetic Field},
  year      = {2011},
  month     = {Jun},
  pages     = {236804},
  volume    = {106},
  doi       = {10.1103/PhysRevLett.106.236804},
  issue     = {23},
  numpages  = {4},
  publisher = {American Physical Society},
  url       = {https://link.aps.org/doi/10.1103/PhysRevLett.106.236804},
}

@InCollection{Liu2023,
title = {Recent developments in fractional {C}hern insulators},
editor = {Tapash Chakraborty},
booktitle = {Encyclopedia of Condensed Matter Physics (Second Edition)},
publisher = {Academic Press},
edition = {Second Edition},
address = {Oxford},
pages = {515-538},
year = {2024},
isbn = {978-0-323-91408-6},
doi = {https://doi.org/10.1016/B978-0-323-90800-9.00136-0},
url = {https://www.sciencedirect.com/science/article/pii/B9780323908009001360},
author = {Zhao Liu and Emil J. Bergholtz}
}

@Article{Anderson,
  author    = {Anderson, P. W.},
  journal   = {Phys. Rev.},
  title     = {Absence of Diffusion in Certain Random Lattices},
  year      = {1958},
  month     = {Mar},
  pages     = {1492--1505},
  volume    = {109},
  doi       = {10.1103/PhysRev.109.1492},
  issue     = {5},
  numpages  = {0},
  publisher = {American Physical Society},
  url       = {https://link.aps.org/doi/10.1103/PhysRev.109.1492},
}

@article{scaling_theory,
  title = {Scaling Theory of Localization: Absence of Quantum Diffusion in Two Dimensions},
  author = {Abrahams, E. and Anderson, P. W. and Licciardello, D. C. and Ramakrishnan, T. V.},
  journal = {Phys. Rev. Lett.},
  volume = {42},
  issue = {10},
  pages = {673--676},
  numpages = {0},
  year = {1979},
  month = {Mar},
  publisher = {American Physical Society},
  doi = {10.1103/PhysRevLett.42.673},
  url = {https://link.aps.org/doi/10.1103/PhysRevLett.42.673}
}

@Article{evers_mirlin,
  author    = {Evers, Ferdinand and Mirlin, Alexander D.},
  journal   = {Rev. Mod. Phys.},
  title     = {Anderson transitions},
  year      = {2008},
  month     = {Oct},
  pages     = {1355--1417},
  volume    = {80},
  doi       = {10.1103/RevModPhys.80.1355},
  issue     = {4},
  numpages  = {0},
  publisher = {American Physical Society},
  url       = {https://link.aps.org/doi/10.1103/RevModPhys.80.1355},
}

@Article{Wegner1979,
  author   = {Wegner, Franz},
  journal  = {Z. Physik B},
  title    = {The mobility edge problem: Continuous symmetry and a conjecture},
  year     = {1979},
  issn     = {1431-584X},
  number   = {3},
  pages    = {207--210},
  volume   = {35},
  doi      = {10.1007/BF01319839},
  refid    = {Wegner1979},
  url      = {https://doi.org/10.1007/BF01319839},
}

@article{qhe_rmp,
  title = {The quantized {H}all effect},
  author = {von Klitzing, Klaus},
  journal = {Rev. Mod. Phys.},
  volume = {58},
  issue = {3},
  pages = {519--531},
  numpages = {0},
  year = {1986},
  month = {Jul},
  publisher = {American Physical Society},
  doi = {10.1103/RevModPhys.58.519},
  url = {https://link.aps.org/doi/10.1103/RevModPhys.58.519}
}

@Article{qi_zhang_rmp,
  author    = {Qi, Xiao-Liang and Zhang, Shou-Cheng},
  journal   = {Rev. Mod. Phys.},
  title     = {Topological insulators and superconductors},
  year      = {2011},
  month     = {Oct},
  pages     = {1057--1110},
  volume    = {83},
  doi       = {10.1103/RevModPhys.83.1057},
  issue     = {4},
  numpages  = {0},
  publisher = {American Physical Society},
  url       = {https://link.aps.org/doi/10.1103/RevModPhys.83.1057},
}

@Article{Hasan_Kane_rmp,
  author    = {Hasan, M. Z. and Kane, C. L.},
  journal   = {Rev. Mod. Phys.},
  title     = {Colloquium: Topological insulators},
  year      = {2010},
  month     = {Nov},
  pages     = {3045--3067},
  volume    = {82},
  doi       = {10.1103/RevModPhys.82.3045},
  issue     = {4},
  numpages  = {0},
  publisher = {American Physical Society},
  url       = {https://link.aps.org/doi/10.1103/RevModPhys.82.3045},
}

@Article{chiu_rmp,
  author    = {Chiu, Ching-Kai and Teo, Jeffrey C. Y. and Schnyder, Andreas P. and Ryu, Shinsei},
  journal   = {Rev. Mod. Phys.},
  title     = {Classification of topological quantum matter with symmetries},
  year      = {2016},
  month     = {Aug},
  pages     = {035005},
  volume    = {88},
  doi       = {10.1103/RevModPhys.88.035005},
  issue     = {3},
  numpages  = {63},
  publisher = {American Physical Society},
  url       = {https://link.aps.org/doi/10.1103/RevModPhys.88.035005},
}

@Article{Laughlin,
  author    = {Laughlin, R. B.},
  journal   = {Phys. Rev. Lett.},
  title     = {Levitation of Extended-State Bands in a Strong Magnetic Field},
  year      = {1984},
  month     = {Jun},
  pages     = {2304--2304},
  volume    = {52},
  doi       = {10.1103/PhysRevLett.52.2304},
  issue     = {25},
  numpages  = {0},
  publisher = {American Physical Society},
  url       = {https://link.aps.org/doi/10.1103/PhysRevLett.52.2304},
}

@Article{Halperin,
  author    = {Halperin, B. I.},
  journal   = {Phys. Rev. B},
  title     = {Quantized {H}all conductance, current-carrying edge states, and the existence of extended states in a two-dimensional disordered potential},
  year      = {1982},
  month     = {Feb},
  pages     = {2185--2190},
  volume    = {25},
  doi       = {10.1103/PhysRevB.25.2185},
  issue     = {4},
  numpages  = {0},
  publisher = {American Physical Society},
  url       = {https://link.aps.org/doi/10.1103/PhysRevB.25.2185},
}

@Article{Onoda_qshe,
  author    = {Onoda, Masaru and Avishai, Yshai and Nagaosa, Naoto},
  journal   = {Phys. Rev. Lett.},
  title     = {Localization in a Quantum Spin {H}all System},
  year      = {2007},
  month     = {Feb},
  pages     = {076802},
  volume    = {98},
  doi       = {10.1103/PhysRevLett.98.076802},
  issue     = {7},
  numpages  = {4},
  publisher = {American Physical Society},
  url       = {https://link.aps.org/doi/10.1103/PhysRevLett.98.076802},
}

@Article{Titum_afai,
  author    = {Titum, Paraj and Berg, Erez and Rudner, Mark S. and Refael, Gil and Lindner, Netanel H.},
  journal   = {Phys. Rev. X},
  title     = {Anomalous {F}loquet-{A}nderson Insulator as a Nonadiabatic Quantized Charge Pump},
  year      = {2016},
  month     = {May},
  pages     = {021013},
  volume    = {6},
  doi       = {10.1103/PhysRevX.6.021013},
  issue     = {2},
  numpages  = {20},
  publisher = {American Physical Society},
  url       = {https://link.aps.org/doi/10.1103/PhysRevX.6.021013},
}

@Article{Prodan_2010,
  author    = {Prodan, Emil and Hughes, Taylor L. and Bernevig, B. Andrei},
  journal   = {Phys. Rev. Lett.},
  title     = {Entanglement Spectrum of a Disordered Topological {C}hern Insulator},
  year      = {2010},
  month     = {Sep},
  pages     = {115501},
  volume    = {105},
  doi       = {10.1103/PhysRevLett.105.115501},
  issue     = {11},
  numpages  = {4},
  publisher = {American Physical Society},
  url       = {https://link.aps.org/doi/10.1103/PhysRevLett.105.115501},
}

@Article{Groth_2014,
  author    = {Christoph W Groth and Michael Wimmer and Anton R Akhmerov and Xavier Waintal},
  journal   = {New J. Phys.},
  title     = {Kwant: a software package for quantum transport},
  year      = {2014},
  month     = {jun},
  number    = {6},
  pages     = {063065},
  volume    = {16},
  doi       = {10.1088/1367-2630/16/6/063065},
  publisher = {IOP Publishing},
  url       = {https://dx.doi.org/10.1088/1367-2630/16/6/063065},
}

@Article{tai_numerics,
  author    = {Li, Jian and Chu, Rui-Lin and Jain, J. K. and Shen, Shun-Qing},
  journal   = {Phys. Rev. Lett.},
  title     = {Topological {A}nderson Insulator},
  year      = {2009},
  month     = {Apr},
  pages     = {136806},
  volume    = {102},
  doi       = {10.1103/PhysRevLett.102.136806},
  issue     = {13},
  numpages  = {4},
  publisher = {American Physical Society},
  url       = {https://link.aps.org/doi/10.1103/PhysRevLett.102.136806},
}

@Article{tai_theory,
  author    = {Groth, C. W. and Wimmer, M. and Akhmerov, A. R. and Tworzyd\l{}o, J. and Beenakker, C. W. J.},
  journal   = {Phys. Rev. Lett.},
  title     = {Theory of the Topological {A}nderson Insulator},
  year      = {2009},
  month     = {Nov},
  pages     = {196805},
  volume    = {103},
  doi       = {10.1103/PhysRevLett.103.196805},
  issue     = {19},
  numpages  = {4},
  publisher = {American Physical Society},
  url       = {https://link.aps.org/doi/10.1103/PhysRevLett.103.196805},
}

@Article{Liu_2020,
  author    = {Liu, Hui and Fulga, Ion Cosma and Asb\'oth, J\'anos K.},
  journal   = {Phys. Rev. Res.},
  title     = {Anomalous levitation and annihilation in {F}loquet topological insulators},
  year      = {2020},
  month     = {Jun},
  pages     = {022048},
  volume    = {2},
  doi       = {10.1103/PhysRevResearch.2.022048},
  issue     = {2},
  numpages  = {5},
  publisher = {American Physical Society},
  url       = {https://link.aps.org/doi/10.1103/PhysRevResearch.2.022048},
}

@article{Huckestein_rmp,
  title = {Scaling theory of the integer quantum {H}all effect},
  author = {Huckestein, Bodo},
  journal = {Rev. Mod. Phys.},
  volume = {67},
  issue = {2},
  pages = {357--396},
  numpages = {0},
  year = {1995},
  month = {Apr},
  publisher = {American Physical Society},
  doi = {10.1103/RevModPhys.67.357},
  url = {https://link.aps.org/doi/10.1103/RevModPhys.67.357}
}

@article{Roy_2016,
  title = {Disordered Chern insulator with a two-step {F}loquet drive},
  author = {Roy, Sthitadhi and Sreejith, G. J.},
  journal = {Phys. Rev. B},
  volume = {94},
  issue = {21},
  pages = {214203},
  numpages = {13},
  year = {2016},
  month = {Dec},
  publisher = {American Physical Society},
  doi = {10.1103/PhysRevB.94.214203},
  url = {https://link.aps.org/doi/10.1103/PhysRevB.94.214203}
}

@article{IQHE_lattice,
  title = {Quantum {H}all--insulator transitions in lattice models with strong disorder},
  author = {Yang, Kun and Bhatt, R. N.},
  journal = {Phys. Rev. B},
  volume = {59},
  issue = {12},
  pages = {8144--8151},
  numpages = {0},
  year = {1999},
  month = {Mar},
  publisher = {American Physical Society},
  doi = {10.1103/PhysRevB.59.8144},
  url = {https://link.aps.org/doi/10.1103/PhysRevB.59.8144}
}

@article{3dTI,
  title = {Effect of strong disorder in a three-dimensional topological insulator: Phase diagram and maps of the ${\mathbb{Z}}_{2}$ invariant},
  author = {Leung, Bryan and Prodan, Emil},
  journal = {Phys. Rev. B},
  volume = {85},
  issue = {20},
  pages = {205136},
  numpages = {10},
  year = {2012},
  month = {May},
  publisher = {American Physical Society},
  doi = {10.1103/PhysRevB.85.205136},
  url = {https://link.aps.org/doi/10.1103/PhysRevB.85.205136}
}

@article{metal_disorder_1,
  title = {Bulk metals with helical surface states},
  author = {Bergman, Doron L. and Refael, Gil},
  journal = {Phys. Rev. B},
  volume = {82},
  issue = {19},
  pages = {195417},
  numpages = {11},
  year = {2010},
  month = {Nov},
  publisher = {American Physical Society},
  doi = {10.1103/PhysRevB.82.195417},
  url = {https://link.aps.org/doi/10.1103/PhysRevB.82.195417}
}

@article{metal_disorder_2,
  title = {Transport through a disordered topological-metal strip},
  author = {Junck, Alexandra and Kim, Kun W. and Bergman, Doron L. and Pereg-Barnea, T. and Refael, Gil},
  journal = {Phys. Rev. B},
  volume = {87},
  issue = {23},
  pages = {235114},
  numpages = {13},
  year = {2013},
  month = {Jun},
  publisher = {American Physical Society},
  doi = {10.1103/PhysRevB.87.235114},
  url = {https://link.aps.org/doi/10.1103/PhysRevB.87.235114}
}

@article{fci,
  title = {Fractional {C}hern Insulator},
  author = {Regnault, N. and Bernevig, B. Andrei},
  journal = {Phys. Rev. X},
  volume = {1},
  issue = {2},
  pages = {021014},
  numpages = {14},
  year = {2011},
  month = {Dec},
  publisher = {American Physical Society},
  doi = {10.1103/PhysRevX.1.021014},
  url = {https://link.aps.org/doi/10.1103/PhysRevX.1.021014}
}

@article{FCI_zero_berry,
  title = {Fractional {C}hern insulators in bands with zero {B}erry curvature},
  author = {Simon, Steven H. and Harper, Fenner and Read, N.},
  journal = {Phys. Rev. B},
  volume = {92},
  issue = {19},
  pages = {195104},
  numpages = {7},
  year = {2015},
  month = {Nov},
  publisher = {American Physical Society},
  doi = {10.1103/PhysRevB.92.195104},
  url = {https://link.aps.org/doi/10.1103/PhysRevB.92.195104}
}

@article{spectral_localizer,
  title = {Local invariants identify topology in metals and gapless systems},
  author = {Cerjan, Alexander and Loring, Terry A.},
  journal = {Phys. Rev. B},
  volume = {106},
  issue = {6},
  pages = {064109},
  numpages = {10},
  year = {2022},
  month = {Aug},
  publisher = {American Physical Society},
  doi = {10.1103/PhysRevB.106.064109},
  url = {https://link.aps.org/doi/10.1103/PhysRevB.106.064109}
}

@article{Liu_prb,
  title = {Mixed higher-order topology: Boundary non-{H}ermitian skin effect induced by a {F}loquet bulk},
  author = {Liu, Hui and Fulga, Ion Cosma},
  journal = {Phys. Rev. B},
  volume = {108},
  issue = {3},
  pages = {035107},
  numpages = {7},
  year = {2023},
  month = {Jul},
  publisher = {American Physical Society},
  doi = {10.1103/PhysRevB.108.035107},
  url = {https://link.aps.org/doi/10.1103/PhysRevB.108.035107}
}

@article{LORING2015383,
title = {K-theory and pseudospectra for topological insulators},
journal = {Annals of Physics},
volume = {356},
pages = {383-416},
year = {2015},
issn = {0003-4916},
doi = {https://doi.org/10.1016/j.aop.2015.02.031},
url = {https://www.sciencedirect.com/science/article/pii/S0003491615000901},
author = {Terry A. Loring}
}

@article{tenfold_symmetry,
  title = {Nonstandard symmetry classes in mesoscopic normal-superconducting hybrid structures},
  author = {Altland, Alexander and Zirnbauer, Martin R.},
  journal = {Phys. Rev. B},
  volume = {55},
  issue = {2},
  pages = {1142--1161},
  numpages = {0},
  year = {1997},
  month = {Jan},
  publisher = {American Physical Society},
  doi = {10.1103/PhysRevB.55.1142},
  url = {https://link.aps.org/doi/10.1103/PhysRevB.55.1142}
}

@article{Carlo_review,
  title = {Random-matrix theory of quantum transport},
  author = {Beenakker, C. W. J.},
  journal = {Rev. Mod. Phys.},
  volume = {69},
  issue = {3},
  pages = {731--808},
  numpages = {0},
  year = {1997},
  month = {Jul},
  publisher = {American Physical Society},
  doi = {10.1103/RevModPhys.69.731},
  url = {https://link.aps.org/doi/10.1103/RevModPhys.69.731}
}

@article{Brouwer,
  title = {Scattering approach to parametric pumping},
  author = {Brouwer, P. W.},
  journal = {Phys. Rev. B},
  volume = {58},
  issue = {16},
  pages = {R10135--R10138},
  numpages = {0},
  year = {1998},
  month = {Oct},
  publisher = {American Physical Society},
  doi = {10.1103/PhysRevB.58.R10135},
  url = {https://link.aps.org/doi/10.1103/PhysRevB.58.R10135}
}

@article{Cosma_2012,
  title = {Scattering theory of topological insulators and superconductors},
  author = {Fulga, I. C. and Hassler, F. and Akhmerov, A. R.},
  journal = {Phys. Rev. B},
  volume = {85},
  issue = {16},
  pages = {165409},
  numpages = {12},
  year = {2012},
  month = {Apr},
  publisher = {American Physical Society},
  doi = {10.1103/PhysRevB.85.165409},
  url = {https://link.aps.org/doi/10.1103/PhysRevB.85.165409}
}

@article{experiment1,
  title = {Broadband Topological Slow Light through Higher Momentum-Space Winding},
  author = {Guglielmon, Jonathan and Rechtsman, Mikael C.},
  journal = {Phys. Rev. Lett.},
  volume = {122},
  issue = {15},
  pages = {153904},
  numpages = {5},
  year = {2019},
  month = {Apr},
  publisher = {American Physical Society},
  doi = {10.1103/PhysRevLett.122.153904},
  url = {https://link.aps.org/doi/10.1103/PhysRevLett.122.153904}
}

@article{experiment3,
  title = {Broadband Topological Slow Light through {B}rillouin Zone Winding},
  author = {Mann, Sander A. and Al\`u, Andrea},
  journal = {Phys. Rev. Lett.},
  volume = {127},
  issue = {12},
  pages = {123601},
  numpages = {6},
  year = {2021},
  month = {Sep},
  publisher = {American Physical Society},
  doi = {10.1103/PhysRevLett.127.123601},
  url = {https://link.aps.org/doi/10.1103/PhysRevLett.127.123601}
}

@article{Schulz-Baldes_2021,
doi = {10.1209/0295-5075/ac1b65},
url = {https://dx.doi.org/10.1209/0295-5075/ac1b65},
year = {2022},
month = {jan},
publisher = {EDP Sciences, IOP Publishing and Società Italiana di Fisica},
volume = {136},
number = {2},
pages = {27001},
author = {Hermann Schulz-Baldes and Tom Stoiber},
title = {Invariants of disordered semimetals via the spectral localizer},
journal = {Europhysics Letters}
}

@article{Schulz-Baldes_2022,
    author = {Schulz-Baldes, Hermann and Stoiber, Tom},
    title = {Spectral localization for semimetals and Callias operators},
    journal = {Journal of Mathematical Physics},
    volume = {64},
    number = {8},
    pages = {081901},
    year = {2023},
    month = {08},
    issn = {0022-2488},
    doi = {10.1063/5.0093983},
    url = {https://doi.org/10.1063/5.0093983}
}

@article{Cosma_localizer,
  title = {Aperiodic Weak Topological Superconductors},
  author = {Fulga, I. C. and Pikulin, D. I. and Loring, T. A.},
  journal = {Phys. Rev. Lett.},
  volume = {116},
  issue = {25},
  pages = {257002},
  numpages = {6},
  year = {2016},
  month = {Jun},
  publisher = {American Physical Society},
  doi = {10.1103/PhysRevLett.116.257002},
  url = {https://link.aps.org/doi/10.1103/PhysRevLett.116.257002}
}

@Article{Braeunlich2010,
  author   = {Bräunlich, G. and Graf, G. M. and Ortelli, G.},
  journal  = {Communications in Mathematical Physics},
  title    = {Equivalence of Topological and Scattering Approaches to Quantum Pumping},
  year     = {2010},
  issn     = {1432-0916},
  number   = {1},
  pages    = {243--259},
  volume   = {295},
  doi      = {10.1007/s00220-009-0983-1},
  refid    = {Bräunlich2010},
  url      = {https://doi.org/10.1007/s00220-009-0983-1},
}

@article{zenodo,
  title        = {Topological fine structure of an energy band},
  author       = {Liu, H. and Fulga, I. C. and Bergholtz, E. J. and Asb{\'o}th, J. K.},
  year         = {2023},
  journal      = {Zenodo},
  pages        = {10359970},
  doi          = {10.5281/zenodo.10359970},
  url          = {https://doi.org/10.5281/zenodo.10359970}
}

@article{10.1063/1.5094300,
    author = {Lozano Viesca, Edgar and Schober, Jonas and Schulz-Baldes, Hermann},
    title = "{Chern numbers as half-signature of the spectral localizer}",
    journal = {Journal of Mathematical Physics},
    volume = {60},
    number = {7},
    pages = {072101},
    year = {2019},
    month = {07},
    issn = {0022-2488},
    doi = {10.1063/1.5094300},
    url = {https://doi.org/10.1063/1.5094300}
}

@article{PhysRevB.109.125132,
  title = {Localization renormalization and quantum Hall systems},
  author = {Andrews, Bartholomew and Reiss, Dominic and Harper, Fenner and Roy, Rahul},
  journal = {Phys. Rev. B},
  volume = {109},
  issue = {12},
  pages = {125132},
  numpages = {10},
  year = {2024},
  month = {Mar},
  publisher = {American Physical Society},
  doi = {10.1103/PhysRevB.109.125132},
  url = {https://link.aps.org/doi/10.1103/PhysRevB.109.125132}
}

@article{PhysRevB.85.075115,
  title = {Topologically protected extended states in disordered quantum spin-Hall systems without time-reversal symmetry},
  author = {Xu, Zhong and Sheng, L. and Xing, D. Y. and Prodan, Emil and Sheng, D. N.},
  journal = {Phys. Rev. B},
  volume = {85},
  issue = {7},
  pages = {075115},
  numpages = {8},
  year = {2012},
  month = {Feb},
  publisher = {American Physical Society},
  doi = {10.1103/PhysRevB.85.075115},
  url = {https://link.aps.org/doi/10.1103/PhysRevB.85.075115}
}

@article{Loring_Hermann_2,
  author  = {Loring, Terry A. and Schulz-Baldes, Hermann},
  title   = {The spectral localizer for even index pairings},
  journal = {Journal of Noncommutative Geometry},
  volume  = {14},
  number  = {1},
  pages   = {1--23},
  year    = {2020},
  doi     = {10.4171/JNCG/357},
  publisher = {EMS Press}
}

@article{experiment2,
  title = {Multiple Brillouin Zone Winding of Topological Chiral Edge States for Slow Light Applications},
  author = {Chen, Fujia and Xue, Haoran and Pan, Yuang and Wang, Maoren and Hu, Yuanhang and Zhang, Li and Chen, Qiaolu and Han, Song and Liu, Gui-geng and Gao, Zhen and Zhou, Peiheng and Yin, Wenyan and Chen, Hongsheng and Zhang, Baile and Yang, Yihao},
  journal = {Phys. Rev. Lett.},
  volume = {132},
  issue = {15},
  pages = {156602},
  numpages = {6},
  year = {2024},
  month = {Apr},
  publisher = {American Physical Society},
  doi = {10.1103/PhysRevLett.132.156602},
  url = {https://link.aps.org/doi/10.1103/PhysRevLett.132.156602}
}

@article{Loring_finite_volume,
 author = {Loring, Terry A. and Schulz-Baldes, Hermann},
 faupublication = {yes},
 journal = {New York Journal of Mathematics},
 pages = {1111 - 1140},
 peerreviewed = {unknown},
 title = {{Finite} volume calculation of {K}-theory invariants},
 doi = {https://nyjm.albany.edu/j/2017/23-48.html},
 volume = {23},
 year = {2017}
}

@article{Selma_localizer,
  title = {Topological zero-modes of the spectral localizer of trivial metals},
  author = {Franca, Selma and Grushin, Adolfo G.},
  journal = {Phys. Rev. B},
  volume = {109},
  issue = {19},
  pages = {195107},
  numpages = {10},
  year = {2024},
  month = {May},
  publisher = {American Physical Society},
  doi = {10.1103/PhysRevB.109.195107},
  url = {https://link.aps.org/doi/10.1103/PhysRevB.109.195107}
}

\end{document}